\documentclass[lettersize,journal]{IEEEtran}
\usepackage{amsmath,amsfonts}
\usepackage{amssymb}
\usepackage{multirow}
\usepackage[ruled,vlined,linesnumbered]{algorithm2e}
\usepackage[table]{xcolor}
\usepackage{array}
\usepackage[caption=false,font=normalsize,labelfont=sf,textfont=sf]{subfig}
\usepackage{textcomp}
\usepackage{stfloats}
\usepackage{siunitx}
\usepackage{url}
\usepackage{verbatim}
\usepackage{graphicx}
\def\BibTeX{{\rm B\kern-.05em{\sc i\kern-.025em b}\kern-.08em
    T\kern-.1667em\lower.7ex\hbox{E}\kern-.125emX}}
\usepackage{balance}
\usepackage{booktabs}
\usepackage{makecell}
\usepackage[normalem]{ulem}
\usepackage{enumitem}
\usepackage{adjustbox}

\newcommand{\emoji}[1]{%
  \raisebox{-0.15em}{\includegraphics[height=1em]{emojis/#1}}%
}

\DeclareMathAlphabet{\pazocal}{OMS}{cmsy}{m}{n}
\usepackage[scr=esstix,cal=boondox]{mathalfa}

\newcommand{\Q}{\pazocal{Q}}

\newcommand{\Loss}{\pazocal{L}}
\newcommand{\f}{f}
\newcommand{\g}{g}

\DeclareMathOperator*{\argmin}{arg\,min}

\newcommand{\rebuttal}[1]{\textcolor{black}{#1}}
\newcommand{\rebuttalvtwo}[1]{\textcolor{black}{#1}}

\begin{document}
\title{HE-NAC: High-Efficiency Neural Audio Coding via Hierarchical and Residual Multi-Band Quantization}
\author{
  Darius P\'etermann, Wootaek Lim,~\IEEEmembership{Member,~IEEE,} Seungkwon Beack, Minje Kim,~\IEEEmembership{Senior Member,~IEEE}\\
\thanks{This work was supported by Electronics and Telecommunications Research Institute (ETRI) grant funded by the Korean government [26ZC1100, Development of Spatial Media Technology and Interaction Technology for Convergence of the Real and Virtual World]. 

Darius P\'etermann is with Indiana University, Bloomington, IN, USA (e-mail: daripete@iu.edu). 
Wootaek Lim and Seungkwon Beack are with Electronics and Telecommunications Research Institute, Daejeon, South Korea (wtlim@etri.re.kr, skbeack@etri.re.kr). Minje Kim (corresponding author) is with the University of Illinois at Urbana-Champaign, Urbana, IL, USA (minje@illinois.edu).
}}

\maketitle

\begin{abstract}

High-fidelity audio coding requires faithful reconstruction of perceptually important high-frequency components, which are often lost due to the aggressive temporal decimation and quantization of neural codecs. State-of-the-art neural audio coding methods rely on adversarial learning to hallucinate these missing components, yielding perceptually plausible but imprecise reconstructions; a limitation for high-fidelity music applications. As a remedy, we propose high-efficiency neural audio coding (HE-NAC), which provides a multi-band extension of the popular residual vector quantization (RVQ) scheme. HE-NAC introduces an effective training mechanism that harmonizes a bitrate-free blind upsampling process, residual learning to compensate it with minimal bitrate, and a multi-head decoder architecture for band-specific reconstruction. As a result, HE-NAC achieves a balanced trade-off, reliably reconstructing the core-band ($\leq$ 3~kHz) component while preserving overall sound fidelity. Subjective listening tests and objective evaluations demonstrate that HE-NAC outperforms both the DAC baseline and MPEG HE-AAC at equal or lower bitrates.
\end{abstract}

\begin{IEEEkeywords}
Neural audio coding, high-efficiency, multiband, residual. 
\end{IEEEkeywords}

\section{Introduction}
Data compression is an essential aspect of information and communication systems nowadays. Its main aim is to achieve a compact alternative representation by eliminating the redundancy of the data present in the signal. This is typically achieved by mapping the raw data samples from their original high precision representation (e.g., single-precision floating points) into a compact discrete representation (e.g., a bitstring). To faithfully reconstruct the original input, a coding system must preserve the attributes that are perceptually or semantically important in its target domain \cite{Herre_2025}. For example, in speech communication, low-bitrate codecs can often maintain \emph{intelligibility} with limited information kept in the bitstream. For general audio signals, however (e.g., music), the required bitrate tends to be much higher since the user's listening experience can be deteriorated even by a subtle perceptual degradation. Scalability to various use cases and bitrates is therefore an important goal in modern speech or audio codecs, for instance as in MPEG unified speech and audio coding (USAC) \cite{usac1, usac2} and Opus \cite{opus}. 

With the advent of deep learning, a novel type of audio coding technology has emerged, namely neural audio coding (NAC) \cite{Kim2024NeuralSA}. NACs introduce data-driven approaches to various parts of a codec, e.g., to decode traditional speech codes in a generative way \cite{KleijnW2018wavenet, ValinJ2019lpcnetcoding}, to learn highly compressed discrete tokens as a foundation model \cite{Baevski2020nips, HsuWN2021ieeeacmaslp, chen2022wavlm}, or as end-to-end autoencoders that take an input waveform, encode it into a compact and abstract representation via a bottleneck, and decode the compressed code back into an approximation of the original signal \cite{KankanahalliS2018icassp, Zeghidour2021soundstream, DefossezA2023encodec, KumarR2023dac}. NACs have made leaps and bounds in many applications, starting with speech generation \cite{Wang2023synth, borsos2023audiolm}, music generation \cite{borsos2023audiolm, rouard2024audio}, speech enhancement \cite{KoizumiY2023miipher, Yang2024Interspeech}, automatic speech recognition (ASR) \cite{Dhawan_2024}, or source separation \cite{Erdogan23_interspeech, bie2025sdcodec, mousavi2025discreteaudiotokenssurvey}. Yet, designing neural \textit{audio} coders that can faithfully reconstruct fine waveform details while maintaining competitive perceptual quality and bitrate efficiency remains an open research challenge.

One of the main sources of NACs’ coding efficiency comes from the encoder’s aggressive temporal decimation (up to 320× in models such as the Descript audio codec (DAC) \cite{KumarR2023dac}), which drastically reduces the data rate at the bottleneck. This decimation, typically implemented through strided convolutional layers, inevitably discards high-frequency details, placing a heavy burden on the decoder to synthesize the missing information during reconstruction, effectively turning the decoding process into a \textit{blind} upsampling task. With most of the energy concentrated in the signal’s \emph{core band}, the decoder's blind upsampling may still manage to reconstruct the signal but with a loss of perceptual quality \cite{EskimezS2019bandwidthex}. Applications such as music coding may substantially suffer from this type of information loss. For instance, excessive temporal decimation leads to a loss of high-frequency information at the bottleneck stage, which places additional burden on the decoder. Given the high standards demanded by high-fidelity music, NACs have therefore struggled to find a proper foothold in real-world music applications.

To circumvent this limitation, recent models have explored generative approaches, such as adversarial learning as proposed in SoundStream \cite{Zeghidour2021soundstream}. In the context of audio reconstruction, the adversarial loss encourages the codec to \textit{generate} the missing information until the discriminator fails to distinguish decoded (fake) signals from input (real) examples. This process pushes the decoder to produce more ``realistic'' and plausible reconstruction that aligns better with the \emph{overall} input distribution, even if that decoded signal deviates from the actual original input. This typically impacts the higher-frequency portions of the signal, which are most affected by the encoding process. As a result, NACs exploiting such a process are often described as ``hallucinating'' high-frequency content (see Fig.~\ref{fig:more_outputs}'s DAC column for some examples). In the context of music, this can be seen as an undesirable effect, where even subtle artifacts can disrupt harmonic structure, timbre, or the listener’s sense of authenticity.

Another crucial technology in modern NAC is the scalable quantization mechanism, exemplified by the residual vector quantization (RVQ) algorithm \cite{Zeghidour2021soundstream}. RVQ leverages the concept of residuals in feature space: a feature vector $\mathbf{x}$ is first approximated by its closest codeword $\mathbf{z}_k$ out of $K$ total codewords in the first codebook, and the difference (residual) $\mathbf{x}-\mathbf{z}_k$ is passed to the next quantizer for further refinement, and so on. This progressive residual modeling naturally orders the series of chosen codewords by their importance, enabling the codec to retain only the top few codewords in low-bitrate scenarios. Intuitively, each residual captures the ``missing'' or fine-grained information not yet represented, a property we later exploit for efficiently modeling high-frequency audio content. \rebuttal{Building upon this principle, the proposed HE-NAC extends the conventional RVQ framework from a single latent representation to a progressive multi-band quantization scheme, where quantization resources can be allocated independently across frequency bands according to the target bitrate and reconstruction requirements.} 

Although the combination of the highly decimated bottleneck features and RVQ is a robust choice, NAC systems with such architecture exhibit progressively more imprecise reconstruction in the higher frequencies. \rebuttal{To address this limitation, we propose HE-NAC, which extends the conventional RVQ coding framework with a hierarchical multi-band residual quantization architecture. Rather than relying on a single bottleneck to represent the entire spectrum, HE-NAC progressively allocates quantization resources across frequency bands, enabling dedicated residual coding of higher-frequency content while preserving the scalability and coding efficiency of RVQ.} We are inspired by the spectral bandwidth replication (SBR) algorithm \cite{Dietz2002SpectralBR}, first introduced with the MPEG-4 high-efficiency advanced audio coding (HE-AAC) \cite{HerreJ2008he-aac}. As in SBR, the proposed HE-NAC employs similar ideas but implemented in a deep learning-based approach: (a) most of the high-frequency information can be inferred from the core-band signal (b) putting more bits in core-band reconstruction is important (c) high-frequency reconstruction requires some code for a more precise reconstruction, rather than relying solely on the decoder's blind and generative upsampling. HE-NAC was designed to consider these principles, resulting in superior sound quality in music audio coding.

In this work, we therefore propose HE-NAC, a multi-band neural audio codec that advances high-fidelity audio coding through the following key contributions:
\begin{itemize}
    \item \rebuttal{We propose HE-NAC, a multi-band extension of the residual vector quantization (RVQ) framework for neural audio coding, enabling progressive band-wise allocation of quantization resources while preserving bitrate scalability.}
    \item HE-NAC's architecture supports a natural harmonization of blind upsampling-based bitrate-free high-frequency reconstruction and dedicated high-frequency coders. Both modules are learned to complement each other, so the potentially inaccurate blind upsampling results are compensated for by the low-bitrate residual coding performed by high-frequency coders. 
    \item We provide a multi-stage training algorithm that provides perceptually meaningful entropy distribution across different subbands, where core band reconstruction quality is emphasized, while the higher-frequency components use just the right amount of code to aid blind upsampling. 
    \item With the multi-stage training, residual learning, and subband coding, HE-NAC provides ``excellent'' quality in MUSHRA-like listening tests \cite{mushra} at 23 kbps, surpassing the baseline DAC at the same bitrate and the commercial implementation of HE-AAC at 24 kbps.
    \item \rebuttal{Finally, while HE-NAC builds upon the progressive multi-band reconstruction paradigm introduced in MB-HARP-Net, it extends this framework into a practical high-fidelity neural audio codec through the integration of RVQ-based latent coding, progressive residual coding with blind upsampling, entropy-aware band-wise bitrate allocation, and a dedicated multi-stage training strategy.}
\end{itemize}

\section{Related Work}

The way NACs handle scalability and signal reconstruction is largely driven by the target domain. For instance, speech codecs often operate under different constraints and requirements than codecs designed for music, mainly because perceptual quality depends differently on different parts of the reconstructed spectrum. These discrepancies across content types directly affect how well a codec scales across bitrates. In this section, we review prior work and group methods by how they approach signal reconstruction, as a function of the target content type and reconstruction strategy. While we acknowledge that there are many possible ways to categorize neural codecs \cite{mousavi2025discreteaudiotokenssurvey}, our chosen organization is intended to clarify what remains missing and to highlight the specific gap that motivates this work.

\paragraph{Plain autoencoder codecs} One way to design an effective NAC is to leverage deep neural network (DNN) architectures. One of the earliest approaches to applying deep convolutional networks for speech compression \cite{KankanahalliS2018icassp} employed scalar quantization and entropy control to regulate the bitrate. This led to a range of NACs that improve perceptual quality over traditional codecs~\cite{BiswasA2020gan-coding,ZhenK2020spl,KorseS2022postgan} or extend to general audio, such as music and ambient sounds~\cite{PetermannD2021harpnet}. While these approaches are promising, their ability to compress general audio signals while preserving high perceptual fidelity is limited to relatively high bitrate cases. In addition, unlike traditional codecs (e.g., AMR-WB~\cite{BessetteB2002amrwb}, Opus, Moving Picture Experts Group's Unified Speech and Audio Coding~\cite{usac1, usac2}), NACs trade increased computational complexity for improved coding gain. These models still struggle to recover fine-grained audio details, for which generative methods are particularly helpful.

\paragraph{Generative Neural Speech Audio Coding} The more homogeneous characteristics of speech signals, compared to general audio signals, have led to pioneering advancements in neural coding. For example, WaveNet~\cite{OordA2016wavenet} learned to autoregressively synthesize speech utterances and can generatively decode a very low-bitrate code~\cite{KleijnW2018wavenet}. However, it remains too computationally heavy for real-time, on-device processing. LPCNet~\cite{ValinJ2019lpcnet} offers a more efficient alternative, though it remains limited to speech coding. Newer generative speech coders~\cite{YangH2024LaDiff,pia2025flowmac} have shown great progress in speech modeling by leveraging diffusion models, but remain too complex due to iterative sampling.

More recently, generative approaches to high-fidelity speech coding have been proposed \cite{binkowski2020adversarial,kong2020hifi,Su2020HiFiGANHD} and have consequently inspired a range of non-speech coding-related tasks. For example, \cite{Zeghidour2021soundstream,DefossezA2023encodec,KumarR2023dac,wu2023audiodec} demonstrated strong potential on high-resolution signals by employing adversarial learning and opened the door to low-bitrate systems capable of handling not only speech but also music and general audio. Generative NACs often use adversarial training with multi-scale discriminators to guide high-frequency reconstruction, as in HiFi-GAN~\cite{kong2020hifi}, which inspired SoundStream~\cite{Zeghidour2021soundstream} and HuBERT-based variants~\cite{polyak2021speech}. However, they often hallucinate high-frequency content: tolerable in speech but detrimental for high-fidelity music, where accuracy and artistic intent are paramount.

\paragraph{Perceptual relevance} Orthogonal to architectural design choices, another line of work focuses on improving reconstruction quality through perceptually motivated training objectives. Since objective loss functions often result in perceptual discrepancies among decoded signals, perceptually-driven loss functions were investigated. These methods showed promising results for NACs, for example, by calibrating the loss with a psychoacoustic weighting scheme~\cite{ZhenK2020spl}, or by integrating a psychoacoustic model (PAM) to measure a perceptual distortion~\cite{ByunJ2022pam}. Such approaches have proven to compete with the well-established MP3 codec, yet at a high bitrate (i.e.,  $>100$~kbps). MDCT-Net introduced predictive modules to the decoder in the modified discrete cosine transform (MDCT) representation \cite{DavidsonG2023mdctnet}: coupled with perceptually weighted methods, it achieved superior quality at 24~kbps on general audio compared to traditional codecs such as Opus.

\paragraph{Multi-Scale and Multi-Band Neural Audio Coding}

An alternate line of work explores a variant of the well-established single-scale RVQ covered so far, by introducing a temporal multi-scale quantization paradigm. SNAC \cite{siuzdak2024snac} proposes to model various audio structures through a hierarchy of quantizers, each working on different temporal resolutions and operating at different time scales. Although demonstrating clear reconstruction improvements over its single-scale quantization counterpart, SNAC still suffers from the high-frequency reconstruction drawback previously outlined. A related line of work is Jukebox~\cite{dhariwal2020jukebox}, which combines a hierarchical Vector Quantized Variational Autoencoder (VQ-VAE) with autoregressive Transformers to model raw audio at multiple temporal scales, enabling the generation of coherent, high-fidelity music with singing over minute-long durations.

Multi-band coding can better capture fine-grained spectral details. Many of these approaches are based on straightforward band-specific coding, e.g., using separate autoencoders per band \cite{roman2023discrete}. On the other hand, the traditional SBR algorithm explicitly models relationships between core and high bands to recover high-frequency content using minimal side information \cite{sbr}. 
From the perspective of machine learning, SBR can be interpreted as an informed version of bandwidth extension (BWE). In other contexts, BWE is typically defined as a ``blind'' process that generates the high-band (HB) component from the core-band (CB) with no side information. This makes it particularly well-suited for machine learning approaches \cite{SunD2013mlsp,EskimezS2019bandwidthex,gupta2019speechbe,su2021bandwidth}. For instance, in the context of NAC, BWE has been used to recover the missing MDCT coefficients based on quantized CB features \cite{shin2019recovery,shin2020enhanced,kim2022recovery}. More recently, \cite{Takahashi2023diffusion} employed multiple diffusion models at different cutoff frequencies, providing hierarchical generation of band-specific signals from low to high. \cite{roman2023discrete} also proposed a band-specific diffusion model that independently processes different frequency bands, reducing cumulative entangled errors. In parallel, MUFFIN \cite{ng2025multiband} introduced a psychoacoustically guided multiband vector quantizer framework, where residually-connected codebooks are dedicated to different spectral regions. MUFFIN demonstrated strong perceptual performance on 24~kHz signals and highlighted the value of treating different spectral regions with distinct modeling and bitrate allocation strategies. Meanwhile, its performance on non-speech audio remains limited.

Collectively, these multi-band approaches to monoaural audio coding have all demonstrated perceptual quality improvements, reinforcing the view that band-structured modeling-whether diffusion-based, residual-VQ–based, or BWE-driven, can provide a more principled path toward efficient high-frequency reconstruction. For high-bandwidth signals such as music, where fine-grained upper-band reconstruction is a key determinant of perceptual quality, these methods tend to saturate: beyond a certain number of codebooks, additional capacity brings diminishing returns for high-frequency reconstruction. We address this scalability gap by proposing a principled, band-aware framework in which the core band still receives the majority of the bitrate, while the remaining budget is explicitly allocated to upper bands according to their perceptual importance. This yields a codec whose scalability is governed by band structure and perceptual weighting, rather than by a single, monolithic bottleneck.

\paragraph{MB-HARP-Net} The multi-band extension of the hyper-autoencoded reconstruction propagation (HARP) network \cite{Petermann2023mbharpnet} explicitly divides the signal into subbands (core-band and high-band), each being encoded and decoded through its own dedicated path. This design enables independent bitrate control and allows the high-band decoder to condition on both its latent code and the upsampled core-band signal, effectively harmonizing neural audio coding with BWE. While its predecessor, HARP-Net \cite{PetermannD2021harpnet}, implicitly learned hierarchical features, MB-HARP-Net enforces band-specific constraints via architectural disentanglement.

\rebuttal{The proposed HE-NAC builds upon this multi-band reconstruction paradigm while extending it into a practical high-fidelity neural audio codec. In particular, unlike MB-HARP-Net, which relies on scalar quantization and entropy regularization, HE-NAC introduces an RVQ-based coding framework with progressive residual quantization, staged blind-upsampling and residual learning, as well as band-wise bitrate allocation through scalable codebook assignment. These additions fundamentally change the coding framework by enabling efficient entropy coding and bitrate scalability while preserving the advantages of progressive multi-band reconstruction.}

\rebuttal{Consequently, MB-HARP-Net primarily provides the architectural foundation for multi-band reconstruction, whereas the contributions of HE-NAC lie in extending this framework into an efficient neural audio codec capable of high-fidelity coding at substantially lower bitrates.}

\section{Neural audio coding preliminaries}
\subsection{Basic autoencoder-based codecs}\label{sec:autoencoder}

At its core, an NAC can be formulated as a deep autoencoder (AE)~\cite{KankanahalliS2018icassp}, learning a compact latent representation of audio for efficient compression. To begin with, the encoder aims to convert the input waveform of $T$ samples $\mathbf{x} \in \mathbb{R}^{T}$ into a sequence of $M_L$-dimensional feature vectors through a series of $L$ neural network blocks:
\begin{equation}
\label{eq:encoding}
\mathbf{H}^{(L)}=[\mathbf{h}_1^{(L)}, \ldots, \mathbf{h}_{N_L}^{(L)}] \leftarrow \pazocal{F}(\mathbf{x}) = \f^{(L)}\circ\cdots\circ f^{(1)}(\mathbf{x}),
\end{equation}
For clarity, we denote by $\mathbf{H}^{(l)}$ the intermediate feature sequence produced after the $l$-th encoder block, with $\mathbf{H}^{(0)} := \mathbf{x}$. Each block transforms its input according to
\begin{equation}
\mathbf{H}^{(l)} \leftarrow f^{(l)}(\mathbf{H}^{(l-1)}), \qquad l = 1,\ldots,L,
\end{equation}
and $\mathbf{H}^{(L)}$ is the final continuous bottleneck representation of dimension $N_L \times M_L$.

For compression, the encoder first reduces the temporal resolution, i.e., $N_L<T$, making strided convolutional layers a natural choice. Second, the nonlinear conversion $\pazocal{F}(\cdot)$ can also prepare the feature vectors for the successive quantization process $\Q(\cdot)$, which replaces each $\mathbf{h}_n^{(L)}\in\mathbb{R}^{M_L}$ with one of $K$ cluster centroids $\mathbf{Z}:=\{\mathbf{z}_k\}_{k=1}^K$:
\begin{equation}
\label{eq:quantization}
\mathbf{h}_n^{(L)}\approx\bar{\mathbf{h}}_n^{(L)}=\mathbf{z}_{n^*}\leftarrow\Q(\mathbf{h}_n^{(L)}), 
\end{equation}
where $n^*=\argmin_{k\in\{1, \ldots, K\}} \pazocal{E}(\mathbf{h}_n^{(L)}, \mathbf{z}_k)$ with a predefined distance metric $\pazocal{E}(\cdot, \cdot)$. We denote the set of selected centroids by $\bar{\mathbf{H}}^{(L)}$, which are the quantized (and dequantized) version of ${\mathbf{H}}^{(L)}$.

Before considering the entropy of the codes (Sec. \ref{sec:entropy_coding}), each $n^*$ index requires $\lceil \log_2 K\rceil$ bits to represent it. Soft-to-hard quantization~\cite{AgustssonE2017softmax} and vector-quantized variational autoencoders (VQ-VAE)~\cite{OordA2017vqvae} are popular choices to achieve this quantization by learning both the codebook $\mathbf{Z}$ and cluster assignments $n^*$, which are also the foundational techniques behind RVQ \cite{Zeghidour2021soundstream, KumarR2023dac}. HE-NAC redefines and extends RVQ into a hierarchical multi-band version (Sec. \ref{sec:proposed}).  

On the decoder side, the quantized feature vectors $\bar{\mathbf{H}}^{(L)}$ are fed to the decoding function $\pazocal{G}(\cdot)$ to recover the original time-domain signal:
\begin{equation}
\label{eq:decoding}
\hat{\mathbf{x}} \leftarrow \pazocal{G}(\bar{\mathbf{H}}^{(L)}) = \g^{(1)}\circ\cdots\circ g^{(L)}(\bar{\mathbf{H}}^{(L)}).
\end{equation}

The decoder's role complements that of the encoder: it seeks to recover the original signal from the information loss introduced by the quantization process Eq.~\eqref{eq:quantization} as well as the temporal decimation from $T$ to $N_L$. The CNN encoder that downsamples the input via strides typically employs upsampling layers on the decoder side. In summary, the autoencoder-based NAC consists of three parts (i) an encoder that transforms the input to an optimal feature representation for coding $\mathbf{H}^{(L)}$, (ii) a quantizer that learns a compact and effective codebook $\mathbf{Z}$ and the assignment of feature vectors to it, and (iii) a decoder that mitigates quantization noise and performs blind upsampling. 
\begin{equation}
\label{eq:nac_reconstruction}
\mathbf{x}\approx\hat{\mathbf{x}} \leftarrow \pazocal{G}\circ\Q\circ\pazocal{F}(\mathbf{x}).
\end{equation}

Note that
the dimension $M_L$ of each of the code vectors is a hyperparameter: as shown in the RVQGAN method in the DAC \cite{KumarR2023dac} model, a large $M_L$ tends to lead to \textit{codebook under-utilization}, which unnecessarily increases the codebook size $K$ and the bitrate unless entropy coding is employed (Sec. \ref{sec:proposed} discusses this issue further).

\subsection{Entropy coding and residual vector quantization}
\label{sec:quant}
\subsubsection{Entropy coding} \label{sec:entropy_coding}
The proposed HE-NAC differs from other NAC models in that it explicitly aims for a lower entropy in its high-frequency codes for an additional lossless compression via entropy coding (e.g., Huffman coding). To this end, the empirical entropy from the frequency of each code $\bar{p}(\mathbf{z}_k)$, i.e., the empirical probability of the centroid $\mathbf{z}_k$ being chosen, can be computed as follows:
\begin{equation}
\label{eq:entropy}
\bar{\pazocal{H}}(\mathbf{Z}):= -\sum_{k=1}^K \bar{p}(\mathbf{z}_k)\log_2 \bar{p}(\mathbf{z}_k),
\end{equation}
where the frequency $\bar{p}(\mathbf{z}_k)$ approximates the unknown probability ${p}(\mathbf{z}_k)$. It is a normalized count of the assignments for a given centroid $\mathbf{z}_k$: $\bar{p}(\mathbf{z}_k)=\frac{1}{I}\sum_{n=1}^I \pazocal{I}(n^*=k)$ with $I$ denoting the total number of latent vectors and $\pazocal{I}(\cdot)$ is an indicator function. This empirical approximation of Shannon’s entropy, i.e., the expected number of bits per codeword, provides a lower bound on the bitrate after applying Huffman coding to $\mathbf{Z}$, by assigning fewer bits to more frequent codewords, and vice versa.

For a sequence of code vectors $\bar{\mathbf{H}}^{(L)}$, before Huffman coding, the bitrate can be calculated by $(R/T) \cdot N_L \cdot \lceil \log_2 K \rceil$, where $R$ is the original input sampling rate (in Hz) and $N_L$ the number of latent frames, and the codebook size $K$. With Huffman coding, the lower bound of the entropy-based bitrate is $(R/T) \cdot N_L \cdot \bar{\pazocal{H}}(\mathbf{Z})$, ideally with $\bar{\pazocal{H}}(\mathbf{Z}) < \lceil \log_2 K \rceil$. Hence, learning a low-entropy code distribution allows more coding gain via lossless Huffman coding. Many existing NAC systems therefore include an entropy regularization term \cite{KankanahalliS2018icassp,PetermannD2021harpnet,YangH2021sanac,zhen2019efficient}, originally proposed in \cite{AgustssonE2017softmax}. While we do not directly regularize the entropy, our proposed hierarchical and residual multi-band quantization encourages a lower entropy in the higher frequency bands, thus achieving a balance between perceptually meaningful reconstruction and a high coding gain.

\subsubsection{Residual vector quantization}
\label{sec:rvq}
More recently, RVQ \cite{Zeghidour2021soundstream} has gained popularity, notably because of its bitrate scalability.  
In RVQ, a series of $J$ vector quantizers \cite{OordA2017vqvae} $[\Q^{(1)},\ldots,\Q^{(J)}]$ are arranged in a multi-stage configuration. Each of the quantizers defines the stage-specific codebook $\mathbf{Z}^{(j)}$. The $j$-th quantization stage $\Q^{(j)}$ works on the residual of the feature vector after subtracting all its preceding stages' codewords from the input. Let the $0$-th stage residual $\mathbf{r}_n^{(0)}$ be one of the floating-point feature vectors, $\mathbf{r}_n^{(0)}=\mathbf{h}_n^{(L)}$. Then, at the $j$-th stage, the residual is defined by:
$\mathbf{r}_n^{(j)}
:= \mathbf{h}_n^{(L)} - \sum_{j'=1}^j
    \Q^{(j')}(\mathbf{r}_n^{(j'-1)})= \mathbf{h}_n^{(L)} - \sum_{j'=1}^j
    \mathbf{z}_{n^*}^{(j')}$,
as $\mathbf{z}_{n^*}^{(j)} \leftarrow \Q^{(j)}(\mathbf{r}_n^{(j-1)})$. In other words, successive stages encode the part of the signal not yet represented by the previous quantizers. After training the RVQ codebooks, the original feature vector can be approximated as the sum of all $J$-stage codewords, $\mathbf{h}_n^{(L)}\approx\sum_{j=1}^J \mathbf{z}_{n^*}^{(j)}$. 

RVQ’s scalability comes from its residual learning scheme, further enhanced by the quantizer-dropout strategy. During training, at each feedforward step, the actual number of quantization stages, $J'$, is sampled from ${1,\ldots,J}$, making the codec more robust to low bitrate cases, i.e., $\mathbf{h}_n^{(L)}\approx\sum_{j=1}^{J'} \mathbf{z}_{n^*}^{(j)}$. 

With RVQ, the raw bitrate increases linearly with the number of active stages $J'$, i.e., $R/T\cdot N^{(L)} \cdot J' \cdot \lceil \log_2 K \rceil$, assuming all codebooks share the same size $K$. For example, DAC explicitly encourages balanced codebook usage by promoting high entropy (i.e., avoiding codebook collapse and underutilization), ensuring that all entries are effectively utilized during training, which in turn improves the efficiency of entropy coding. In contrast, the entropy-coded bitrate depends on the entropy of each codebook: $ R/T\cdot N^{(L)} \cdot \sum_{j=1}^{J'}\bar{\pazocal{H}}(\mathbf{Z}^{(j)})$. For example, if some codebooks exhibit lower entropy due to the residual structure, entropy coding becomes more effective. 

In HE-NAC, we propose to harmonize the RVQ concept with a multi-band setting, maximizing the benefit of RVQ through the interplay between residual learning and high-frequency reconstruction.

\section{The Proposed HE-NAC model}
\label{sec:proposed}

\begin{figure}
    \centering
    \includegraphics[width=1.\linewidth]{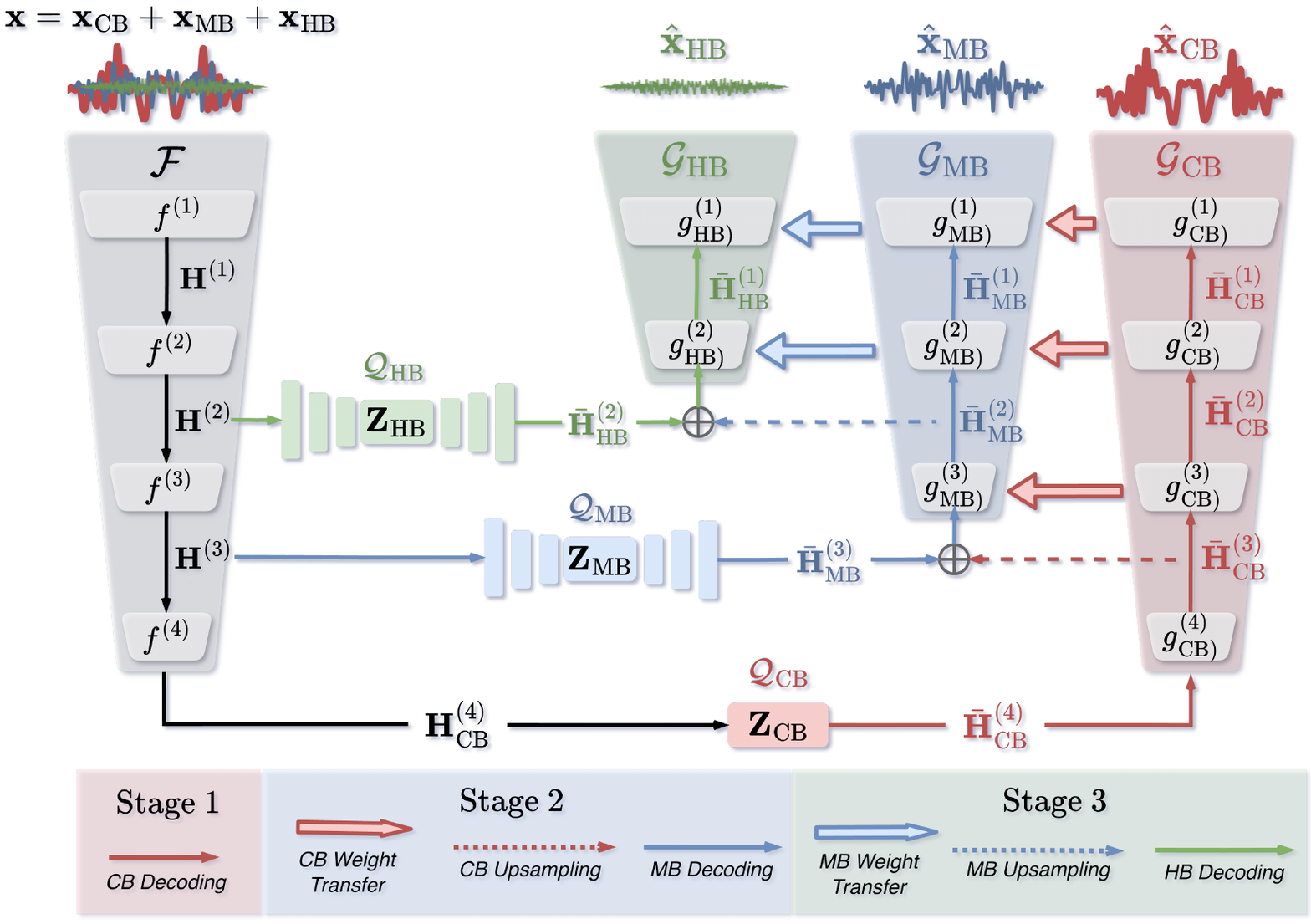}
    \caption{Overview of the proposed HE-NAC system.}
    \label{fig:main}
\end{figure}

HE-NAC extends the DAC backbone into the MB-HARP-Net architecture. It inherits HARP-Net’s advantages (e.g., skip-connected encoder–decoder design) while gaining subband-specific modeling flexibility from MB-HARP-Net. The advantages of this design are:

\begin{enumerate}
    \item An improved band-specific reconstruction quality via dedicated loss defined with band-passed filtered targets;
    \item A combination of blind upsampling and residual coding to achieve balanced modeling of core and high-frequency bands;
    \item A multi-stage training procedure that enables hierarchical, residual, and multi-band quantization for efficient and scalable encoding.
\end{enumerate}
Fig. \ref{fig:main} illustrates the HE-NAC training process.

\subsection{MB-HARP-Net extension of DAC}

HE-NAC's CNN encoder produces $L$ progressively decimated latent layers, which are processed by multiple decoding heads individually. The $l$-th subband representation $\mathbf{H}^{(l)}$ (see Sec.~\ref{sec:autoencoder}) corresponds to the latent feature map produced at depth $l$. Without loss of generality, we assume that $L=4$ with layers $l \in \{2,3,4\}$ representing the high-, mid-, and core-band components, respectively. This hierarchy reflects the progressively larger stride factors in deeper encoder blocks, with $l=4$ producing the most heavily downsampled representation. $N_l$ denotes the number of feature vectors obtained after downsampling from the initial $T$ time-domain samples. Hence, $T>N_1>N_2>N_3>N_4$. 

At an intermediate layer $l<4$, the latent vectors $\mathbf{H}^{(l)}$ are also sent to the decoder side using the \textit{skip-AE} paradigm. When $l=2$ (high band, HB) and $l=3$ (mid-band, MB), the skip-AEs operate as follows:

\begin{equation}\label{eq:skipae_hb}
\bar{\mathbf{H}}_\text{HB}^{(2)}
\leftarrow
\pazocal{V}_\text{HB}
\circ
\pazocal{Q}_\text{HB}
\circ
\pazocal{U}_\text{HB}
\bigl(\mathbf{H}^{(2)}\bigr),
\end{equation}

\begin{equation}\label{eq:skipae_mb}
\bar{\mathbf{H}}_\text{MB}^{(3)}
\leftarrow
\pazocal{V}_\text{MB}
\circ
\pazocal{Q}_\text{MB}
\circ
\pazocal{U}_\text{MB}
\bigl(\mathbf{H}^{(3)}\bigr).
\end{equation}
where $\pazocal{U}_\text{HB}$, $\pazocal{V}_\text{HB}$, and $\pazocal{Q}_\text{HB}$ denote the high-band skip-AE encoder, decoder, and quantizer, respectively. For the high-band codebooks $\{\mathbf{Z}_\text{HB}^{(j)}\}_{j=1}^{J_\text{HB}}$, the value of $J_\text{HB}$ determines the number of RVQ codebooks used in the HB skip-AE. Analogous components are defined for the mid-band.

The core-band (CB) representation, $\mathbf{H}^{(4)}$, does not employ a skip-AE, but a direct quantization step:
\begin{equation}
\bar{\mathbf{H}}_\text{CB}^{(4)} = \pazocal{Q}_\text{CB}\bigl(\mathbf{H}^{(4)}\bigr),
\end{equation}
where $\pazocal{Q}_\text{CB}$ denotes the core-band quantizer.

Assigning specific architectural components to individual spectral bands enables flexible and scalable modeling. We consider this a key strength of HE-NAC. 

HE-NAC employs a multi-head decoding architecture that expands the single-head decoder defined in Sec.~\ref{sec:autoencoder} into $\pazocal{G}_\text{HB}(\cdot)$,  $\pazocal{G}_\text{MB}(\cdot)$, and  $\pazocal{G}_\text{CB}(\cdot)$, each of which takes the output of the corresponding band-specific skip-AE. $\pazocal{G}_\text{CB}(\cdot)$ is equivalent to the ordinary decoder $\pazocal{G}(\cdot)$:
\begin{equation}\label{eq:multi-head-decoding}
\begin{aligned}
\hat{\mathbf{x}}_\text{HB}\!\leftarrow\!\pazocal{G}_\text{HB}(\bar{\mathbf{H}}_\text{HB}^{(2)}), ~~\hat{\mathbf{x}}_\text{MB}\!\leftarrow\!\pazocal{G}_\text{MB}(\bar{\mathbf{H}}_\text{MB}^{(3)}),~~
\hat{\mathbf{x}}_\text{CB}\!\leftarrow\!\pazocal{G}_\text{CB}(\bar{\mathbf{H}}_\text{CB}^{(4)}),
\end{aligned}
\end{equation}
where each band-specific decoder has a reduced number of blocks according to the level of abstraction of their respective input features, i.e., $\pazocal{G}_\text{HB}(\cdot)=\g_\text{HB}^{(1)}\circ \g_\text{HB}^{(2)}(\cdot)$ and $\pazocal{G}_\text{MB}(\cdot)=\g_\text{MB}^{(1)}\circ \g_\text{MB}^{(2)}\circ \g_\text{MB}^{(3)}(\cdot)$.

Each band-specific model is optimized to reconstruct its corresponding high-pass filtered target signal. However, controlling the bitrate for each band is challenging, particularly at higher layers (small $l$), where the data rate $\frac{N_l}{T/R}$ is high due to the finer temporal resolution. To address this, HE-NAC introduces a hierarchically structured quantization scheme that harmonizes the upsampling path (Sec.~\ref{sec:residual_learning}) with multi-stage training (Sec.~\ref{sec:multi-stage}).

\subsection{Harmonization of blind upsampling and per-band reconstruction}\label{sec:residual_learning}
Efficient reconstruction of high-frequency content is key to successful high-fidelity audio coding. HE-NAC leverages the blind upsampling capability of the DAC backbone model (i.e., reconstructing upper bands from the bottleneck code) while also incorporating the band-specific code provided by the MB and HB skip-AEs. To this end, we redefine Eq.~\eqref{eq:multi-head-decoding} using the upsampled intermediate features:
\begin{equation}
\begin{aligned}
    \hat{\mathbf{x}}_\text{HB} &\leftarrow \pazocal{G}_\text{HB}\!\left(\bar{\mathbf{H}}_\text{HB}^{(2)} + \bar{\mathbf{H}}_\text{MB}^{(2)}\right), \\
    \hat{\mathbf{x}}_\text{MB} &\leftarrow \pazocal{G}_\text{MB}\!\left(\bar{\mathbf{H}}_\text{MB}^{(3)} + \bar{\mathbf{H}}_\text{CB}^{(3)}\right), \quad
    \hat{\mathbf{x}}_\text{CB} \leftarrow \pazocal{G}_\text{CB}\!\left(\bar{\mathbf{H}}_\text{CB}^{(4)}\right).
\end{aligned}
\end{equation}
Here, the CB decoding path $\hat{\mathbf{x}}_\text{CB} \leftarrow \pazocal{G}_\text{CB}(\bar{\mathbf{H}}_\text{CB}^{(4)})$ still relies solely on DAC's blind upsampling (rightmost path in Fig.~\ref{fig:main}). Conversely, in the MB and HB decoding paths, the intermediate upsampled reconstructions from the corresponding lower-band decoder (dotted lines in Fig.~\ref{fig:main}) are added to the feature space of the higher bands. For MB decoding, for example,  
\[
\bar{\mathbf{H}}_\text{MB}^{(2)} \leftarrow 
    \g_\text{MB}^{(3)}\!\left(\bar{\mathbf{H}}_\text{MB}^{(3)} + \bar{\mathbf{H}}_\text{CB}^{(3)}\right),
\quad
\bar{\mathbf{H}}_\text{CB}^{(3)} \leftarrow 
    \g_\text{CB}^{(4)}(\bar{\mathbf{H}}_\text{CB}^{(4)}).
\]

Note that $\bar{\mathbf{H}}_\text{CB}^{(3)}$ and its succeeding tensors in the CB decoding path are relying solely on blind upsampling to recover the full-band signal. The MB decoding, however, merges the blind upsampling capability, inherited from the CB path, and the MB-specific coding, by combining $\bar{\mathbf{H}}_\text{CB}^{(3)}$ and $\bar{\mathbf{H}}_\text{MB}^{(3)}$ as its input. Similarly, the MB decoding path still blindly upsamples to recover the HB region; for this reason, the HB decoding path compensates it by blending $\bar{\mathbf{H}}_\text{MB}^{(2)}$ and $\bar{\mathbf{H}}_\text{HB}^{(2)}$. Hence, the contributions of ${\mathbf{Z}}_\text{CB}$ to $\hat{\mathbf{x}}_\text{MB}$ and $\hat{\mathbf{x}}_\text{HB}$, ${\mathbf{Z}}_\text{MB}$ to $\hat{\mathbf{x}}_\text{HB}$ are \textit{bitrate-free} (i.e., obtained via blind upsampling), while ${\mathbf{Z}}_\text{CB}$, ${\mathbf{Z}}_\text{MB}$, and ${\mathbf{Z}}_\text{HB}$ are meant to contribute to their corresponding subbands' reconstruction, respectively.

\subsection{Multi-stage training}\label{sec:multi-stage}

A multi-stage training strategy provides a hierarchical and residual learning structure for HE-NAC, ensuring the desired bitrate allocation to different subbands. The overall procedure described in this section is summarized in Alg.~\ref{alg:he-nac}. 
Concretely, this setup enforces a top–down residual-learning structure: each stage models only the information not captured in its previous stage. Consequently, the learned representations are expected to satisfy 
$\sum_{j=1}^{J_\text{CB}} \bar{\pazocal{H}}(\mathbf{Z}_\text{CB}^{(j)}) >
 \sum_{j=1}^{J_\text{MB}} \bar{\pazocal{H}}(\mathbf{Z}_\text{MB}^{(j)}) >
 \sum_{j=1}^{J_\text{HB}} \bar{\pazocal{H}}(\mathbf{Z}_\text{HB}^{(j)})$,
since higher stages encode progressively smaller residuals. For clarity, from now on we assume to be working with 32~kHz input signals (i.e., with a Nyquist frequency of 16~kHz).

\subsubsection{CB-only upsampling: the lowest bitrate codec}
\label{sec:cb-only}
\begin{figure*}
    \centering
    \includegraphics[width=1.\textwidth]{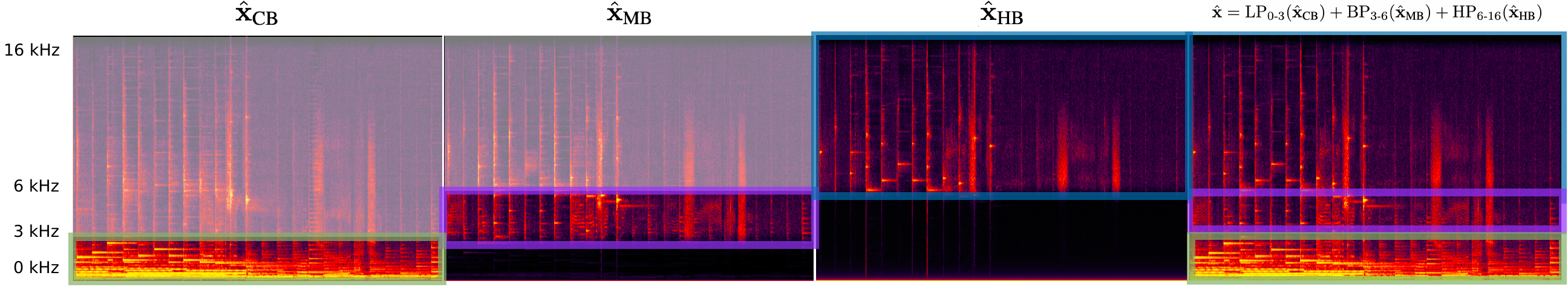}
    \caption{\rebuttal{Overall reconstruction flow of the proposed three-stage training paradigm. Gray segments denote components used only during training: they provide blind-upsampled, bitrate-free priors to the next stage, but are omitted (filtered out) at inference.}}
    \label{fig:reconstruction}
\end{figure*}

The CB decoder is first trained along with the encoder that produces the most temporally and spectrally decimated features:
\begin{equation}
    \hat{\mathbf{x}}_\text{CB}\!\leftarrow\!\pazocal{G}_\text{CB}(\bar{\mathbf{H}}_\text{CB}^{(4)}),~\bar{\mathbf{H}}_\text{CB}^{(4)}\!\leftarrow\!\pazocal{Q}_\text{CB}(\mathbf{H}^{(4)}),~ \mathbf{H}^{(4)}\!\leftarrow\!\pazocal{F}(\mathbf{x}),\!\!\!\!\!\!
\end{equation}
whose output is compared to the original 0-16~kHz full-band signal using a loss function $\Loss({\mathbf{x}}||\hat{\mathbf{x}}_\text{CB})$. It is important to emphasize that, although the CB codebooks primarily represent core-band (~0-3~kHz) information due to their strong temporal and frequency decimation, the CB decoder is nevertheless trained to reconstruct the \emph{entire} full-band waveform, i.e., the loss $\Loss(\mathbf{x} \| \hat{\mathbf{x}}_\text{CB})$ compares full-band signals. Consequently, frequencies above the core band are produced purely through the blind-upsampling mechanism of $\pazocal{G}_\text{CB}$, without any band-specific codebooks. 
\rebuttal{The chosen frequency boundaries reflect the reconstruction characteristics observed in modern RVQ-based neural codecs rather than an assumed optimal spectral decomposition. In practice, single-bottleneck codecs already reconstruct the lower-frequency region (i.e., 0--3 kHz) with high fidelity, while reconstruction errors become progressively more pronounced at higher frequencies. We therefore preserve the strengths of the baseline codec in the core band and dedicate subsequent stages to progressively refining the spectral regions where additional coding is most beneficial. This design philosophy is also consistent with traditional high-efficiency codecs such as HE-AAC v1, which employ different coding strategies for lower and higher frequency regions instead of treating the spectrum uniformly.}
\rebuttal{This design is intentional: rather than restricting each stage to reconstruct only its designated frequency band, the decoder is trained to reconstruct both its target band and all higher-frequency regions. Consequently, the CB, MB, and HB decoders respectively predict the 0-16~kHz, 3-16~kHz, and 6-16~kHz portions of the spectrum. This allows each stage to provide a blind-upsampling prior for the subsequent residual stage, while only the frequency band assigned to that stage contributes to the final reconstructed signal (Fig.~\ref{fig:reconstruction}).}
This path corresponds to many autoencoder-based systems that rely heavily on adversarial loss for their blind upsampling path \cite{Zeghidour2021soundstream, DefossezA2023encodec, KumarR2023dac}. While adversarially trained blind upsampling can produce perceptually plausible high frequencies, it does not preserve the exact waveform structure required for high-fidelity coding. This gap motivates adding explicit residual stages. That being said, once trained, this CB-only stage can act as a standalone codec operating at the lowest bitrate, directly comparable to the DAC baseline. 

Before introducing the MB codebooks, the entire CB path $(\pazocal{F}, \pazocal{Q}_\text{CB}, \pazocal{G}_\text{CB})$ is \emph{frozen}. This prevents the model from reshaping the CB representation when optimizing the MB loss and ensures that the MB stage learns a genuine residual representation on top of a fixed CB-only operating point. Freezing CB is therefore essential for preserving entropy monotonicity and for maintaining consistent bitrate scalability across stages. The following stages (MB and HB) will therefore focus exclusively on predicting the residual information missing from this CB reconstruction in the upper-band frequencies ($> 3$~kHz).

\subsubsection{Decoding of MB codes for residual learning in the mid-bitrate codec}

The MB stage follows a structure analogous to the CB stage, with one key difference: 
although the MB codebooks primarily target the mid-band region ($\sim$3-6~kHz), the 
decoder is trained to reconstruct the entire upper-band range (3-16~kHz). To 
ensure that MB focuses exclusively on information not captured by the CB-only 
decoder, we first apply a high-pass filter that removes the 0-3~kHz band from the target signal, producing $\mathbf{x}_{\text{MB}}$. The MB decoder, therefore, is optimized using a band-limited loss (3-16~kHz). In practice, we decimate the MB code space to a level where it is sufficient to represent the mid-band content ($\sim$3-6~kHz), while the remaining high frequencies 
($\sim$6-16~kHz) are reconstructed through the model’s generative capability and are 
further refined during the subsequent HB stage. This mirrors the logic used for 
CB: each stage is trained to approximate its target band using only the 
information available in its own quantized representation, while still being 
responsible for reconstructing its upper spectral region. \rebuttal{Consequently, the frequency interpretation in HE-NAC is imposed by the decoder supervision and staged training, rather than by explicitly partitioning the latent representations themselves into frequency-domain subbands.} We now describe how the MB skip-AE and decoder layers are integrated into this residual path.

The second training stage introduces the MB codebooks $\{\mathbf{Z}_\text{MB}^{(j)}\}_{j=1}^{J_\text{MB}}$ into the decoding process to complement the CB-only path. The weights of $g^{(3)}_\text{MB}(\cdot)$, $g^{(2)}_\text{MB}(\cdot)$ and $g^{(1)}_\text{MB}(\cdot)$ are initialized from those of $g^{(3)}_\text{CB}(\cdot)$, $g^{(2)}_\text{CB}(\cdot)$ and $g^{(1)}_\text{CB}(\cdot)$, respectively, to ensure that the MB decoder initially behaves like the CB decoder. That way, any improvement over the CB-only reconstruction must come from the residual contribution of $\bar{\mathbf{H}}_\text{MB}^{(3)}$. This is essential for enforcing a low entropy in the MB codes, which can otherwise explode due to their higher data rate. 

As described in Sec. \ref{sec:residual_learning}, the input to the MB decoder is the sum of two components: the bitrate-free upsampled feature $\bar{\mathbf{H}}_\text{CB}^{(3)} \leftarrow g^{(4)}_\text{CB}(\bar{\mathbf{H}}_\text{CB}^{(4)})$, and the bit-consuming reconstruction from the MB skip-AE, $\bar{\mathbf{H}}_\text{MB}^{(3)}$, both of which share the same temporal resolution.

Fine-tuning the MB decoder layers allows the contribution from the MB codes to be gradually incorporated. The same loss function, $\Loss(\mathbf{x}_\text{MB} \| \hat{\mathbf{x}}_\text{MB})$, is used, but this time it compares the output against the high-pass filtered target $\mathbf{x}_\text{MB}$, which also includes HB components. %

\subsubsection{Decoding of HB codes: the highest bitrate codec}
Similar ideas are applied to higher bands. Once both $\pazocal{G}_\text{CB}(\cdot)$ and $\pazocal{G}_\text{MB}(\cdot)$ are pretrained sequentially, the CB and MB decoding paths are frozen. The parameters of $g^{(2)}_\text{MB}$ and $g^{(1)}_\text{MB}$ are transferred to HB-specific decoding layers, respectively, $g^{(2)}_\text{HB}$ and $g^{(1)}_\text{HB}$, and further fine-tuned. This ensures that the HB stage learns a true residual representation on top of the previously fixed CB and MB reconstructions.

\begin{algorithm}[th]
\label{alg:he-nac}
\SetAlgoLined
\SetKw{Frozen}{Frozen \emoji{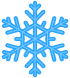}}
\SetKw{Trainable}{Trainable \emoji{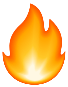}}
\SetKwInOut{Input}{Input}
\SetKwInOut{Output}{Output}
\SetKw{Frozen}{Frozen \emoji{snowflake}}
\SetKw{Trainable}{Trainable \emoji{fire}}

\Input{Training dataset $\pazocal{D} = \{\mathbf{x}\}$}
\Output{Trained HE-NAC model parameters}

\textbf{Initialization:} Randomly initialize $\pazocal{F}$, 
$\pazocal{G}_\text{CB}, \pazocal{G}_\text{MB}, \pazocal{G}_\text{HB}$, 
and codebooks $\pazocal{Q}_\text{CB}, \pazocal{Q}_\text{MB}, \pazocal{Q}_\text{HB}$.

\SetKwProg{Stage}{Stage}{}{end}

\Stage{Stage 1: Core-band (CB) low-bitrate codec}{
    \Trainable $\pazocal{F}, \pazocal{Q}_\text{CB}, \pazocal{G}_\text{CB}$ \;
    Target: Full-band signal $\mathbf{x}$ \;
    Loss: $\Loss(\mathbf{x} \| \hat{\mathbf{x}}_\text{CB})$ \;
    Result: standalone CB codec + generative approximation for the MB/HB frequency ranges \;
}
\vspace{0.3em}

\Stage{Stage 2: Mid-band (MB) residual codec}{
    \Frozen $\pazocal{F}, \pazocal{Q}_\text{CB}, \pazocal{G}_\text{CB}$ \;
    \Trainable $\pazocal{U}_\text{MB}, \pazocal{V}_\text{MB}, \pazocal{Q}_\text{MB}, \pazocal{G}_\text{MB}$ \;
    Input: 
    \begin{itemize}
        \item Bitrate-free upsampled CB features \\ $\bar{\mathbf{H}}_\text{CB}^{(3)} \gets g_\text{CB}^{(4)}(\bar{\mathbf{H}}_\text{CB}^{(4)})$
        \item MB skip-AE features $\bar{\mathbf{H}}_\text{MB}^{(3)}$
    \end{itemize}
    Target: High-pass filtered signal $\mathbf{x}_\text{MB}$ (3-16~kHz) \;
    Loss: $\Loss(\mathbf{x}_\text{MB} \| \hat{\mathbf{x}}_\text{MB})$ \;
    Initialization: MB decoder weights copied from CB decoder \;
    Result: The MB decoder with improved MB reconstruction with residual coding + HB generative approximation \;
}
\vspace{0.3em}

\Stage{Stage 3: High-band (HB) residual codec}{
    \Frozen $\pazocal{F}, \pazocal{Q}_\text{CB}, \pazocal{G}_\text{CB}, \pazocal{U}_\text{MB}, \pazocal{V}_\text{MB}, \pazocal{Q}_\text{MB}, \pazocal{G}_\text{MB}$ \;
    \Trainable $\pazocal{U}_\text{HB}, \pazocal{V}_\text{HB}, \pazocal{Q}_\text{HB}, \pazocal{G}_\text{HB}$ \;
    Input:
    \begin{itemize}
        \item Upsampled MB+CB features \\ $\bar{\mathbf{H}}_\text{MB}^{(2)} \gets g_\text{MB}^{(3)}(\bar{\mathbf{H}}_\text{MB}^{(3)} + \bar{\mathbf{H}}_\text{CB}^{(3)})$
        \item HB skip-AE features $\bar{\mathbf{H}}_\text{HB}^{(2)}$
    \end{itemize}
    Target: high-pass filtered signal $\mathbf{x}_\text{HB}$ (6-16~kHz) \;
    Loss: $\Loss(\mathbf{x}_\text{HB} \| \hat{\mathbf{x}}_\text{HB})$ \;
    Initialization: HB decoder weights copied from MB decoder \;
    Result: The HB decoder with improved HB reconstruction \;
}
\caption{Overview of the proposed multi-stage training paradigm for HE-NAC.}
\end{algorithm}

The HB decoder receives a high-pass filtered target signal $\mathbf{x}_\text{HB}$ corresponding to the highest frequency band. Its input is the sum of the HB-specific skip-AE feature $\bar{\mathbf{H}}_\text{HB}^{(2)}$ and the upsampled MB contribution $\bar{\mathbf{H}}_\text{MB}^{(2)}$. The decoding result $\hat{\mathbf{x}}_\text{HB}$ is optimized with a band-limited loss over the HB range (6-16~kHz), ensuring that only residual high-frequency information not captured by CB or MB is encoded. Since $\pazocal{G}_\text{MB}(\cdot)$ has already been fully trained with the objective of reconstructing the mid-to-high frequency range (3-16~kHz), $\hat{\mathbf{x}}_\text{HB}$ recovers the residual in the highest frequency band with a minimal bitrate.

\subsubsection{Full-band reconstruction and bitrate scalability}
At inference time, the final full-band reconstruction is obtained by combining the outputs of the active stages in a band-wise manner. When only CB codes are used, the output is given directly by the CB decoder, $\hat{\mathbf{x}} = \hat{\mathbf{x}}_\text{CB}$, which provides a full-band signal through blind upsampling.
When MB codes are enabled, the reconstruction replaces the upper-band content of the CB output. Specifically, the 0–3~kHz band is taken from $\hat{\mathbf{x}}_\text{CB}$, while the 3–16~kHz band is taken from $\hat{\mathbf{x}}_\text{MB}$.
When both MB and HB codes are enabled, each decoder contributes to its corresponding frequency range. The final signal is obtained as
\[
\hat{\mathbf{x}} = 
\mathrm{LP}_{0\text{-}3}\!\left(\hat{\mathbf{x}}_\text{CB}\right)
+ \mathrm{BP}_{3\text{-}6}\!\left(\hat{\mathbf{x}}_\text{MB}\right)
+ \mathrm{HP}_{6\text{-}16}\!\left(\hat{\mathbf{x}}_\text{HB}\right)
\]
where $\text{LP}$, $\text{BP}$, and $\text{HP}$ denote ideal low-pass, band-pass, and high-pass filtering, respectively.
This band-wise replacement ensures that each stage contributes only to the frequency region it is responsible for, yielding a scalable bitrate codec in which higher stages progressively refine the reconstruction without altering the lower-band content.

\section{Experiments}

From now on, we use the following nomenclature to denote the various configurations of HE-NAC: HE-NAC$_{(J_\text{CB}, J_\text{MB}, J_\text{HB})}$.

\subsection{Data}
All models, including the DAC baseline, are trained using the free music archive (FMA) dataset \cite{DefferrardM2017fma}. All signals were originally MP3-encoded at 320~kbps, and were downsampled from 44.1~kHz to 32~kHz. Although not as ideal as using uncompressed signals, we found that the high bitrate used in FMA is sufficient given the relatively low target bitrates of our codecs. Most importantly, we note that many commercial codecs natively downsample their input signals. The HE-AAC baseline we use, for example, cuts off the frequency at around 16~kHz for 24 kbps, and progressively reduces this cutoff at lower bitrates. We partition the \texttt{Large} FMA subset into training, validation, and test sets, comprising approximately 86k, 20k, and 1k songs, respectively.

\subsection{Training Regime}

Similar to the original DAC implementation, we use the AdamW optimizer \cite{loshchilov2019adamw} with an initial learning rate of $1\times 10^{-4}$. We set our effective batch size to 32 with a chunk size of 0.38 seconds and train our systems over $300k$ iterations.%

\subsection{Training Objective}

Our training objective is inspired by the original DAC formulation and consists of a combination of band-specific frequency reconstruction $\pazocal{L}_{\text{rec}}$, codebook learning $\pazocal{L}_{\text{cb}}$, and adversarial losses $\pazocal{L}_{\text{adv}}$. As a reminder, the reconstruction of each band is optimized in incremental stages. In concrete terms, in the first stage of HE-NAC, namely HE-NAC$_{(J_\text{CB},0,0)}$, the full-band input and output, $\mathbf{x}$ and $\hat{\mathbf{x}}_\text{CB}$, respectively, are accounted for. In the second stage, the high-pass filtered $\mathbf{x}_\text{MB}$ (at the cutoff frequency ($\sim$3~kHz)) and $\hat{\mathbf{x}}_\text{MB}\leftarrow \text{HE-NAC}_{(J_\text{CB},J_\text{MB},0)}$ are compared. Lastly, in the final stage, $\mathbf{x}_\text{HB}$ (6~kHz and above) and $\hat{\mathbf{x}}_\text{HB}\leftarrow \text{HE-NAC}_{(J_\text{CB},J_\text{MB},J_\text{HB})}$ are compared.

\subsubsection{{Adversarial Loss}}

Our band-specific adversarial optimization strategy is defined as follows:
\begin{equation}
\pazocal{L}_{\text{adv}}^{(b)}(\hat{\mathbf{x}}_b) = \frac{1}{N_{\text{disc}}} \sum_n \text{max}\bigl(0, 1 - D_n(\hat{\mathbf{x}}_b)\bigr)
\label{eq:loss_gen}
\end{equation}
where $D_n$ denotes the $n$-th discriminator with  $N_{\text{disc}}$ being the total number of discriminators. $b$ denotes one of the subbands: $b \in \{\text{CB}, \text{MB}, \text{HB}\}$. The use of discriminators operating at different time-frequency resolutions, originally introduced in \cite{DefossezA2023encodec}, has proven to be particularly effective at capturing different structures in audio signals.

The adversarial loss term can be enforced by computing the mean absolute difference in the features from the internal layers of the discriminators between the generated and original target audio samples. Formally, this process can be defined as:
\begin{equation}
\Loss_{\text{feat}}(\mathbf{x}_b, \hat{\mathbf{x}}_b) = \frac{1}{N_{\text{disc}}L} \sum_{n=1}^{N_{\text{disc}}}\sum_{l=1}^{L} \frac{\| D_{n}^{l}(\mathbf{x}_b) - D_{n}^{l}(\hat{\mathbf{x}}_b) \|_1}{\mathrm{mean}\left(\|D_{n}^{l}(\mathbf{x}_b) \|_1\right)},
    \label{eq:loss_feat}
\end{equation}
where $L$ denotes the number of layers contained in each of the $N_{\text{disc}}$ discriminators. Finally, the discriminators are trained on the following hinge loss:
\begin{equation}
\begin{split}
\Loss_{\text{disc}}(\mathbf{x}_b, \hat{\mathbf{x}}_b) = \frac{1}{N_{\text{disc}}} \sum_{n=1}^{N_{\text{disc}}} &\max\bigl(0, 1 - D_n(\mathbf{x}_b)\bigr) + \\
&\max\bigl(0, 1 + D_n(\hat{\mathbf{x}}_b)\bigr)
    \label{eq:loss_disc}
\end{split}
\end{equation}

\subsubsection{Reconstruction Loss}
In a similar fashion, reconstruction optimization is handled per band according to the specific training stage. We use an $L_1$ loss on Mel spectrogram representations computed at seven different time and frequency resolutions, i.e., window sizes of $[32, 64, 128, 256, 512, 1024, 2048]$ and corresponding numbers of Mel subbands $[5, 10, 20, 40, 80, 160, 320]$. Hop sizes are fixed to $25\%$ of the window sizes. 

Importantly, the Mel filterbanks are adapted to the target frequency range of each stage by setting band-dependent minimum and maximum frequencies $(f_{\min}, f_{\max})$. Specifically, we use $(0, 3\,\text{kHz})$ for CB, $(3, 6\,\text{kHz})$ for MB, and $(6, 16\,\text{kHz})$ for HB, ensuring that the loss focuses only on the spectral region modeled at each stage.

\begin{equation}
\pazocal{L}_{\text{rec}} = \sum_{i=1}^{7} 
\left\| \text{Mel}^{(i)}_{[f_{\min}, f_{\max}]}(x) 
- \text{Mel}^{(i)}_{[f_{\min}, f_{\max}]}(\hat{x}) \right\|_1
\end{equation}
where $\text{Mel}^{(i)}_{[f_{\min}, f_{\max}]}$ denotes the Mel spectrogram computed using the $i$-th time-frequency resolution and a filterbank restricted to the frequency range $[f_{\min}, f_{\max}]$.

\subsection{Model Details}

\rebuttalvtwo{All reported bitrates are computed after entropy coding of quantized latent indices (Huffman), i.e., as effective bitrate. This differs from standard DAC reporting, which commonly uses the nominal fixed-rate budget from frame rate and number of codebooks (pre-entropy). In HE-NAC, entropy coding is not a detached post-processing step: the proposed multi-stage residual design (especially for mid/high bands) explicitly shapes index statistics toward lower entropy, so effective bitrate reduction is part of the codec behavior itself. To clarify this distinction, we report both nominal pre-entropy rate and entropy-coded effective rate.}

We detail below the architectural hyperparameters used for our proposed model. The backbone encoder–decoder is based on the DAC architecture, with an encoder composed of four convolutional layers with downsampling rates of $[2, 2, 5, 20]$, and a decoder with corresponding upsampling rates of $[20, 5, 2, 2]$. The bottleneck's dimension is $1,536$, which is projected again to an $8$ dimensional space before RVQ, and uses $31$ codebooks, each of size $1,024$. The codebooks are trained with the codebook dropout strategy as well. \rebuttalvtwo{This enables bitrate-scalable operation by varying the number of codebooks allocated to the CB, MB, and HB branches. Across the configurations evaluated in this work, HE-NAC operates over an effective bitrate range of approximately 13-23~kbps after Huffman entropy coding.}

For HE-NAC, we incorporate two additional skip AEs, which are used to encode intermediate features from the main encoder using lightweight DAC modules. The first skip autoencoder compresses mid-band features with a single-layer encoder and decoder of stride $5$, which essentially replaces the baseline's last stride of 20 (i.e., $[2, 2, 5, 5]$ downsampling rates). The MB skip AE trains $4$ codebooks of size $4,096$ with codebook dropouts. The HB skip autoencoder targets high-band features by hijacking the features after the second encoder layer and applying its own stride rate of 16 (i.e., $[2, 2, 16]$ downsampling rates). It uses $2$ codebooks of size $16,384$ with codebook dropouts. Note that when MB and/or HB skip AEs are active, the CB pathway (corresponding to the baseline DAC) operates with a reduced number of codebooks (e.g., 16 instead of 31), with the remaining bitrate budget assigned to the MB and/or HB codebooks. Due to both the smaller number of codebooks and the residual nature of the MB and HB representations, these higher-band codes exhibit lower entropy. This mitigates the trade-off between bitrate allocation and the perceptual importance of different frequency ranges.

\begin{table}
\centering
\footnotesize
{\color{black}
\setlength{\tabcolsep}{3.5pt}
\caption{DAC vs HE-NAC at the $\sim$23 kbps operating point. Raw rate is pre-entropy (10 bits/token, $K=1024$).}
\label{tab:model_comparison}
\resizebox{\columnwidth}{!}{
\begin{tabular}{llccccc}
\toprule
\rowcolor{gray!20}
\textbf{Model} & \textbf{Band} & \textbf{Stride} & \textbf{CB} & \textbf{FR (Hz)} & \textbf{Ent. kbps} & \textbf{Raw kbps} \\
\midrule
DAC$_{(31)}$ & Full & 400 & 31 & 80 & 23.0 & 24.8 \\
\midrule
\multirow{4}{*}{HE-NAC$_{(16,4,2)}$}
& Core & 400 & 16 & 80  & 12.0 & 12.8 \\
& Mid  & 100 & 4  & 320 & 6.0  & 12.8 \\
& High & 64  & 2  & 500 & 5.0  & 10.0 \\
\cmidrule(lr){6-7}
& Total& --  & -- & --  & 23.0 & 35.6 \\
\bottomrule
\end{tabular}
}
}
\end{table}

We incorporate the following model configurations in our comparative study. HE-NAC's codebook dropout during training allows us to choose an adequate number of codebooks for each band to ensure the desired bitrates after Huffman coding across different model configurations. For clarity, we report the empirically measured bitrates obtained after entropy coding, and use their rounded values (to the nearest kbps) in the following.
\begin{itemize}[leftmargin=*, itemindent=0pt, itemsep=0pt] 
\item HE-NAC$_{(16, 4, 2)}$: Our proposed 3-band model with a high bitrate configuration (22.984 kbps), attributing $J_\text{CB}=16$, $J_\text{MB}=4$, and $J_\text{HB}=2$ codebooks for CB, MB, and HB, respectively. 
\item HE-NAC$_{(16, 4, 0)}$: Our low-bitrate model with 18.278 kbps, which does not use the HB code, thus relying on blind upsampling to reconstruct the highest frequency range (6-16 kHz).
\item DAC$_{(31)}$: The original DAC baseline with $J=31$ codebooks, whose bitrate is 23.013 kbps. Architecturally, it is equivalent to the CB code-only model (Sec. \ref{sec:cb-only}). It is missing both the MB and HB codes.
\item HE-AAC v1: The AAC codec enhanced with SBR for efficient high-frequency reconstruction at low bitrates. We use the commercial implementation in Adobe Audition\textsuperscript{\textregistered}. We set its bitrate to be 24 kbps, which is the lowest bitrate covering $\sim$32 kHz in Adobe Audition\textsuperscript{\textregistered}.
\end{itemize}
The cutoff frequencies for MB and HB high-pass filtering are at 3~kHz and 6~kHz, respectively.

\begin{table}
\caption{Objective results across various DAC model with sizes matching our proposed model. Metrics are computed on full-band input/output}
\centering
\resizebox{\columnwidth}{!}{
\begin{tabular}{l c c c c c}
\toprule
\rowcolor{gray!20}
\textbf{Models} & \textbf{\#Params} &
\makecell{\textbf{Mel-}\\\textbf{distance} $\downarrow$} &
\makecell{\textbf{STFT}\\\textbf{-distance} $\downarrow$} &
\makecell{\textbf{VISQOL}\\$\uparrow$} &
\makecell{\textbf{SI-SDR}\\$\uparrow$} \\
\addlinespace[2pt]
\toprule
DAC$_{(31)}$ & 85M &  0.42 & 1.68 & 4.33 & 15.6 \\
DAC$_{(31)}$ & 127M &  0.41 & 1.67 & 4.36 & 15.7 \\
DAC$_{(31)}$ & 220M &  0.41 & 1.67 & 4.35 & 15.7 \\[1.5ex]
\hline 
\\[-0.5ex]  %
HE-NAC$_{(16, 4, 2)}$ & 220M & 0.52  &  1.76  & 4.40  & 13.7 \\
\bottomrule
\end{tabular}
\label{tab:more_objective_results}
}
\end{table}

Due to the proposed multi-decoding paradigm, our model entails more parameters than the original DAC. To ensure a fair comparison, we use a parameter-matched DAC variant for our baseline. To isolate the effect of model size itself, we include a small table (see Table~\ref{tab:more_objective_results}) depicting the signal-to-distortion ratio (SI-SDR) \cite{LeRouxJL2018sisdr} reconstruction performance of DAC as a function of its model size. \rebuttal{We retrain DAC using the same sampling rate, target bitrate, and training protocol as HE-NAC to isolate the architectural differences between the two codecs. Although official DAC checkpoints are publicly available, they target different operating points (16, 24, or 44.1~kHz, with the 44.1~kHz models operating only at 8 or 16~kbps), making direct comparisons difficult to interpret due to the simultaneous change in sampling rate, reconstruction bandwidth, and bitrate.}

Commercial codecs must handle the trade-off between the target bitrate and the coverage of the frequency bandwidth. Our HE-AAC v1 baseline chooses to downsample the input signal to 29 kHz instead of trying to cover the higher frequencies with its small bit budget at 24 kbps. Our DAC baseline and HE-NAC models chose to use 32 kHz as their full-band reconstruction target, which provides reasonable fidelity considering the bitrate range of 18-24 kbps.

\rebuttal{Relatively few publicly available neural codecs operate under the same high-fidelity setting targeted in this work, namely near-transparent audio coding at approximately 20--24~kbps with a 32~kHz reconstruction bandwidth. Several recent codecs therefore address different operating regimes rather than serving as direct competitors.}

Since our goal is to achieve a reasonably high fidelity with the 32 kHz target sample rate, some recent neural codecs, such as MUFFIN \cite{ng2025multiband} and BSCodec \cite{wang2026bscodec}, are not directly comparable to our methods as they operate at 24 kHz. SNAC \cite{siuzdak2024snac} was learned from 32 kHz signals, but its target bitrate is too low (1.9 kbps) for a fair comparison to our high-fidelity codecs at higher bitrates. \rebuttal{For these reasons, we focus our evaluation on DAC as a representative neural audio codec under matched operating conditions and HE-AAC v1 as a strong commercial baseline in this bitrate range.}

\section{Results}

We present a comprehensive evaluation of the proposed approach, including subjective listening tests, objective metrics, and targeted ablation studies. Together, these experiments assess perceptual quality, reconstruction fidelity, and the contribution of each component of our proposed model.

\subsection{Subjective listening tests}\label{sec:mushra}

\begin{figure}
\centering
    \includegraphics[width=0.8\linewidth]{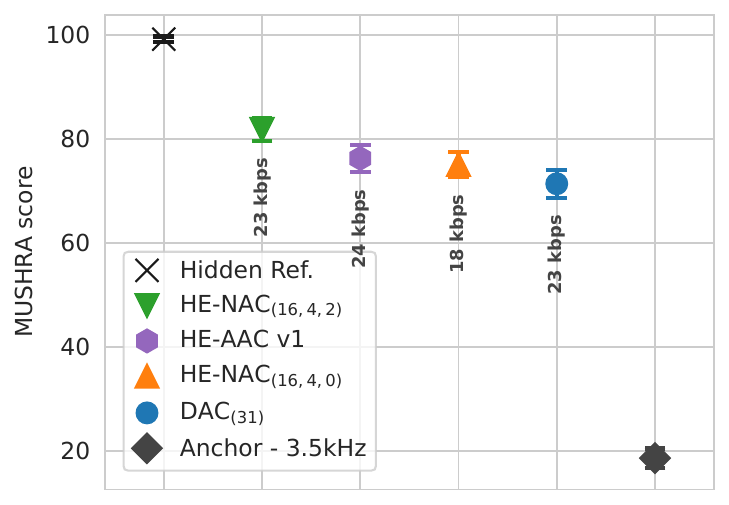}\vspace{-0.01in}
  \caption{Mean MUSHRA scores for each condition with 95\% confidence intervals for the mean. Nominal bitrates (kbps) are indicated under the coded systems.
  }

  \label{fig:mushra}    
\end{figure}

We first conduct MUSHRA-style \cite{mushra} tests to evaluate the perceptual sound quality of the baseline and our proposed codec. We recruited 16 audio experts, but excluded one of them who did not meet the post-screening guideline. Our subjective listening test is implemented by webMUSHRA \cite{Schoeffler2018webmushra} and consists of 12 randomly chosen 10-second excerpts from the FMA test set. Subjects used high-quality headphones. Each trial includes the hidden reference, a low-pass filtered anchor with a cutoff at 3.5 kHz, and the four systems in comparison: HE-NAC$_{(16, 4, 2)}$ (23 kbps), HE-NAC$_{(16, 4, 0)}$ (18 kbps), DAC$_{(31)}$ (23 kbps), and HE-AAC v1 (24 kbps).

Fig.~\ref{fig:mushra} depicts the results of the MUSHRA test. We observe that our proposed 23 kbps system HE-NAC$_{(16, 4, 2)}$ performs best, significantly surpassing both DAC at 23 kbps and HE-AAC v1 at 24 kbps. The fact that our high-bitrate mode HE-NAC$_{(16, 4, 2)}$ performs significantly better than the low-bitrate mode HE-NAC$_{(16, 4, 0)}$ indicates that the bits allocated adequately to the HB codes can significantly improve the HB reconstruction in the 6-16 kHz range. We also note that the MB-only model HE-NAC$_{(16, 4, 0)}$ operating at 18 kbps outperforms the 23 kbps DAC baseline. It empirically proves the proposed staged multiband neural coding is effective: re-allocating the bits for the latter half of the codebooks in DAC (from codebook 17 to 31) to the four MB codebooks ($\{\mathbf{Z}_\text{MB}^{(j)}\}_{j=1}^{J_\text{MB}=4}$) results in a perceptually better reconstruction. Moreover, the same MB-only configuration reaches scores very close to HE-AAC v1 at 24 kbps despite operating at a 6 kbps lower bitrate.

\subsection{Objective evaluation}

\begin{table}
\caption{Objective results across individual subbands and full bands. Lower is better for Mel/STFT distance, higher is better for VISQOL/SI-SDR.}
\centering

\setlength{\tabcolsep}{3pt}
\begin{tabular}{@{\hskip 0pt}l c c c c c@{}}
\toprule
kbps & Models & \makecell{Mel\\distance $\downarrow$} &
\makecell{STFT\\distance $\downarrow$} &
\makecell{VISQOL$\uparrow$} &
\makecell{SI-SDR$\uparrow$} \\
\toprule

\rowcolor{gray!20} \multicolumn{6}{c}{\textbf{Core-Band (0kHz - 3kHz)}} \\
\midrule

\multirow{2}{*}{\hspace{3pt}18}  
& DAC$_{(24)}$          & \bfseries 0.49 & \bfseries 0.32 & -- & \bfseries 18.0 \\
& HE-NAC$_{(16,4,0)}$   & 0.62 & 0.41 & -- & 15.4 \\

\arrayrulecolor{black!15}\midrule
\arrayrulecolor{black}

\multirow{2}{*}{\hspace{3pt}23}
& DAC$_{(31)}$          & \bfseries 0.37 & \bfseries 0.24 & -- & \bfseries 20.6 \\
& HE-NAC$_{(16,4,2)}$   & 0.62 & 0.41 & -- & 15.4 \\

\midrule
\rowcolor{gray!20} \multicolumn{6}{c}{\textbf{Mid-Band (3kHz - 6kHz)}} \\
\midrule

\multirow{2}{*}{\hspace{3pt}18}
& DAC$_{(24)}$          & 1.05 & 0.36 & -- & -1.4 \\
& HE-NAC$_{(16,4,0)}$   & \bfseries 0.83 & \bfseries 0.29 & -- & \bfseries 3.7 \\

\arrayrulecolor{black!15}\midrule
\arrayrulecolor{black}

\multirow{2}{*}{\hspace{3pt}23}
& DAC$_{(31)}$          & 0.96 & 0.33 & -- & 0.7 \\
& HE-NAC$_{(16,4,2)}$   & \bfseries 0.84 & \bfseries 0.29 & -- & \bfseries 3.6 \\

\midrule
\rowcolor{gray!20} \multicolumn{6}{c}{\textbf{High-Band (6kHz - 12kHz)}} \\
\midrule

\multirow{2}{*}{\hspace{3pt}18}
& DAC$_{(24)}$          & 0.91 & 1.08 & -- & \bfseries -32.6 \\
& HE-NAC$_{(16,4,0)}$   & \bfseries 0.90 & 1.08 & -- & -35.4 \\

\arrayrulecolor{black!15}\midrule
\arrayrulecolor{black}

\multirow{2}{*}{\hspace{3pt}23}
& DAC$_{(31)}$          & 0.90 & 1.08 & -- & -30.7 \\
& HE-NAC$_{(16,4,2)}$   & \bfseries 0.83 & \bfseries 1.04 & -- & \bfseries -5.5 \\

\midrule
\rowcolor{gray!20} \multicolumn{6}{c}{\textbf{Full-Band}} \\
\midrule

\multirow{2}{*}{\hspace{3pt}18}
& DAC$_{(24)}$          & \bfseries 0.49 & \bfseries 1.79 & \bfseries 4.32 & \bfseries 14.2 \\
& HE-NAC$_{(16,4,0)}$   & 0.53 & 1.80 & 4.31  & 13.0 \\

\arrayrulecolor{black!15}\midrule
\arrayrulecolor{black}

\multirow{2}{*}{\hspace{3pt}23}
& DAC$_{(31)}$          & \bfseries 0.41 & \bfseries 1.67 & 4.35 & \bfseries 15.7 \\
& HE-NAC$_{(16,4,2)}$   & 0.52 & 1.76 & \bfseries 4.40  & 13.7 \\

\bottomrule
\end{tabular}
\label{tab:objective_results}
\end{table}

\begin{figure}[t] %
\centering
\includegraphics[width=1.\linewidth]{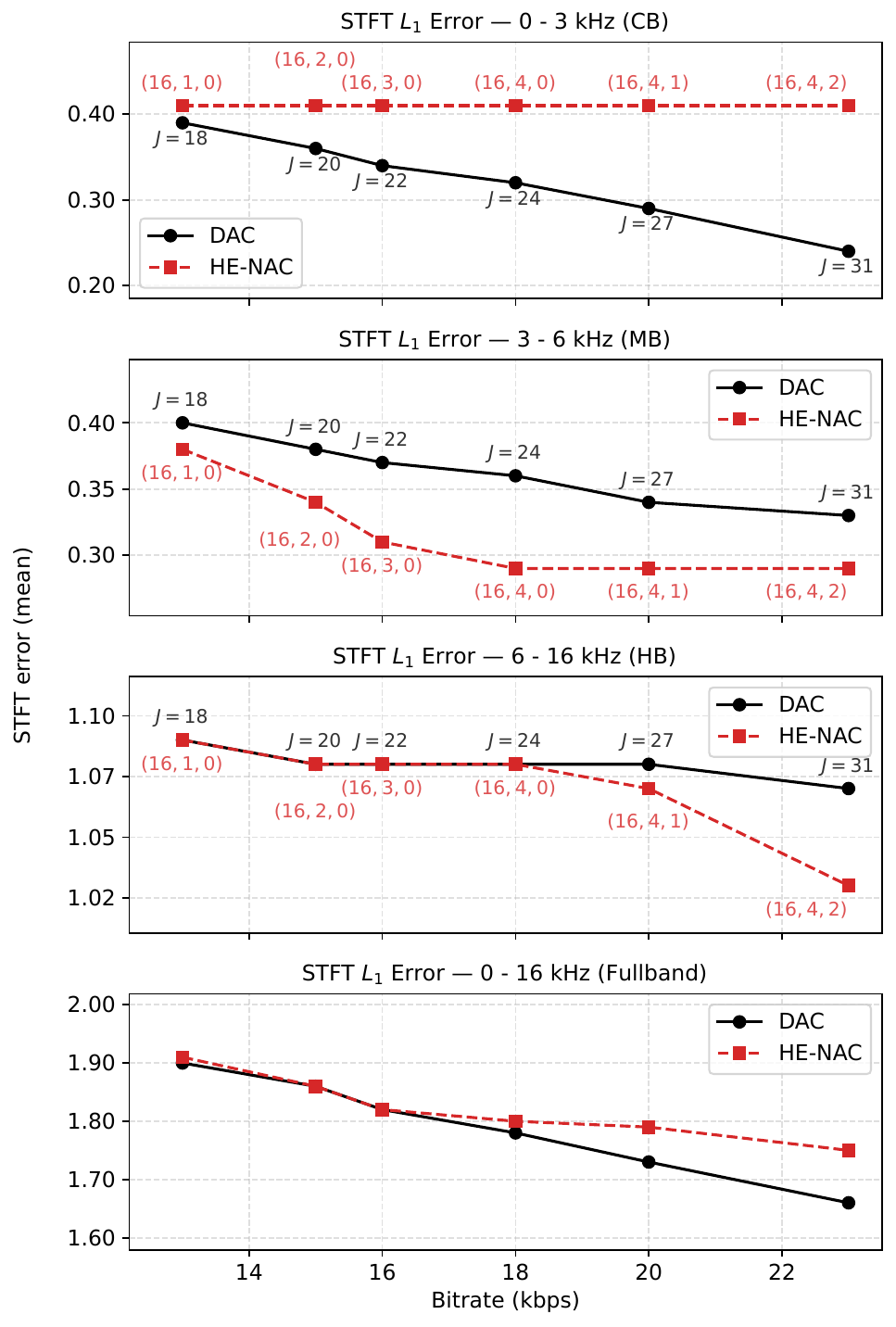}
\caption{Rate–distortion (STFT error) curves shown per frequency band (0-3,
3-6, 6-16 kHz, and full-band) over the 13-23 kbps range. DAC (baseline)
scales its bitrate by controlling the number of codebooks, ranging from 18 to
31, defined by its single bottleneck. HE-NAC (proposed) keeps 16 codebooks
in its CB, but then gradually increases the MB codebooks from 1 to 4, as
well as the HB codebooks from 0 to 2. \rebuttalvtwo{Each DAC datapoint is
annotated with its total number of codebooks, $J$, whereas each HE-NAC
datapoint is annotated with the complete branch-specific allocation
$(J_{\mathrm{CB}},J_{\mathrm{MB}},J_{\mathrm{HB}})$. The third entry,
$J_{\mathrm{HB}}$, therefore denotes the number of codebooks allocated to
the HB branch.} \rebuttal{When $J_{\mathrm{HB}}=0$,} no dedicated HB
residual is transmitted and the 6-16 kHz band is reconstructed solely
through blind upsampling from the CB and MB stages. \rebuttalvtwo{Increasing
$J_{\mathrm{HB}}$ to 1 and 2} progressively enables dedicated HB residual
coding, explaining the sharp reduction in HB distortion.}
\label{fig:distortion_rate}
\end{figure}

Table~\ref{tab:objective_results} summarizes the objective comparison results on a per-band basis for spectral distance metrics (Mel and STFT) as well as the SI-SDR signal-level metric. The full-band results incorporate VISQOL \cite{visqol_v3_2020} as well. We first note that the $\text{DAC}$ baselines (24 and 31 codebooks) perform best in core-band reconstruction, suggesting its representational power is concentrated at low frequencies, although it has little effect on perceptual quality as shown in Sec. \ref{sec:mushra}. In the mid-band, $\text{HE-NAC}_{(16,  4, 0)}$ and $\text{HE-NAC}_{(16, 4, 2)}$ both demonstrate strong performance, surpassing their baseline by $5.04$ and $2.90$-dB SI-SDR, respectively. On the HB performance, $\text{HE-NAC}_{(16, 4, 2)}$ performs best, reducing all three metrics by a large margin. 
These improvements translate to a more balanced full-band perceptual VISQOL metric (on top of MUSHRA), with highest VISQOL score of $4.40$.

Despite large SI-SDR gains in the high-frequency (e.g., $+25$ dB for $\text{HE-NAC}_{(16, 4, 2)}$), full-band SI-SDR shows minimal change despite its positive perceptual contribution. As well-known in the literature (e.g., \cite{HuY2008speech_quality}), this suggests that the SNR-variant metrics are dominated by low-frequency content and may under-represent improvements in high-band fidelity in the context of full-band quality assessment, necessitating subjective listening tests as shown in Sec. \ref{sec:mushra}.
\rebuttal{This behavior is expected, as HE-NAC was not designed to improve the already strong core-band reconstruction of DAC. Instead, the proposed architecture allocates additional coding capacity to progressively refine the mid- and high-frequency components, which are typically the most challenging to reconstruct accurately with conventional single-bottleneck neural codecs.}

Fig.~\ref{fig:distortion_rate} echoes the same observation made above through STFT error-rate performance curves. Since in our HE-NAC setup $J_\text{CB}$ is left unchanged, the baseline DAC's performance in the CB frequencies gradually increases as DAC's CB codebook size increases (the top subfigure). Meanwhile, since the HE-NAC increases its bitrate by increasing the MB and HB codebooks, i.e., $J_\text{MB}$ and $J_\text{HB}$, they naturally do not contribute to the reconstruction in the CB frequencies. However, the MB-only performance (second to the top), with a contribution as small as $J_\text{MB}=1$, already shows considerable reconstruction improvement over the DAC baseline, with the gap widening as $J_\text{MB}$ is increased up to $4$. Lastly, for the HB-only reconstruction (the third figure), we start seeing a considerable drop in reconstruction error as soon as $J_\text{HB}$ is introduced (20 kbps). 

Overall, combined with strong subjective scores, our results suggest that HE-NAC achieves effective bit allocation across subbands, enabling more perceptually relevant mid- and high-frequency reconstruction.

\begin{figure}
    \centering
        \includegraphics[width=1.0\linewidth]{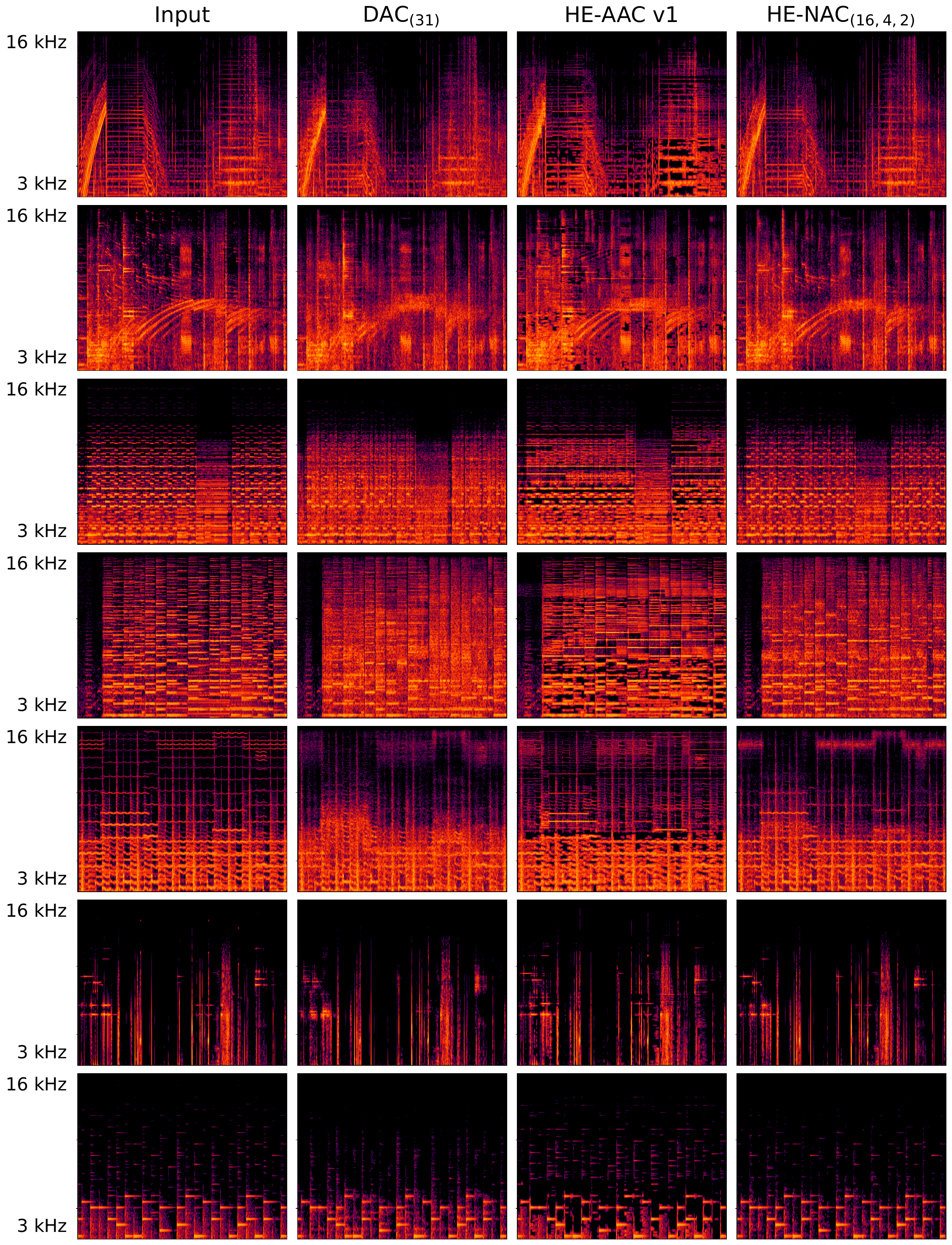}
    \caption{Comparisons of spectrograms (3kHz-16kHz) across several audio examples and models. Our model showcase the best reconstruction in that band considering the additional coding artifacts added in HE-AAC.} 
    \label{fig:more_outputs}
\end{figure}

Fig.~\ref{fig:more_outputs} presents additional examples of mid-to-high band reconstruction. We include our $\text{HE-NAC}_{(16, 4, 2)}$, DAC$_{(31)}$ (both at 23 kbps), and HE-AAC v1 (with SBR) at 24 kbps.

Overall, both HE-AAC and HE-NAC$_{(16, 4, 2)}$ demonstrate strong HB reconstruction capabilities, while HE-AAC's results show spectral holes, which contribute to the perceptual degradation. We also observe that DAC$_{(31)}$ performs reasonably well, although the reconstructions are overall smoothed, missing details in the harmonic structures, especially when these happen to be complex. We claim that these are precisely the scenarios where our method excels, as it leverages explicit high-frequency information rather than attempting to infer it from lower-band cues alone.

\subsection{Model Size Comparison}

We present additional objective metrics for the DAC models considered in this study. As noted earlier, our proposed HE-NAC model is larger than the DAC baseline. To isolate the impact of model size from architectural improvements, we compare DAC variants whose parameter counts match those of $\text{HE-NAC}_{(16, 4, 0)}$ and $\text{HE-NAC}_{(16, 4, 2)}$. Table~\ref{tab:more_objective_results} summarizes the results.

We observe that performance remains nearly constant across the DAC models with increasing parameter counts, especially on frequency-sensitive metrics, such as Mel- and STFT-distance. This suggests that simply increasing the model capacity beyond the original DAC model does not yield significant performance gains: instead, the quantization schemes and bit allocation across different codebooks are the bottleneck of coding performance. In contrast, our HE-NAC model achieves competitive or better performance in HB-sensitive metrics (e.g., VISQOL, STFT-distance) despite architectural differences, indicating that its improvements are not solely attributable to the capacity of the model. We again emphasize that our model outperforms the DAC baseline in subjective tests (Sec. \ref{sec:mushra}), although it may underperform on metrics biased toward low-frequency information, such as Mel-distance or SI-SDR.

\subsection{Residual Representation}

\begin{table}[t]
\caption{SI-SDR (dB) and GOMPSNR across bands (3, 6, 16~kHz, and full-band) over 13--23 kbps for DAC and HE-NAC. VISQOL scores are included for the full-band evaluation. For HE-NAC, the tuples $(J_\mathrm{CB},J_\mathrm{MB},J_\mathrm{HB})$ denote the number of RVQ codebooks allocated to the core-, mid-, and high-band branches, respectively. Since DAC employs a single RVQ bottleneck, only the total number of codebooks is applicable.}
\centering
\sisetup{
  detect-weight,
  mode=text,
  table-number-alignment=center,
  round-mode=places,
  round-precision=1
}

\setlength{\tabcolsep}{1pt}  %
\label{tab:sisdr_visqol_results}

\resizebox{\columnwidth}{!}{
\begin{tabular}{l l @{\,\hspace{4pt}\,} *{6}{S}}
\toprule

\multicolumn{2}{c}{} &
\multicolumn{6}{c}{\textbf{kbps}} \\
 &  & \textbf{13} & \textbf{15} & \textbf{16} &
       \textbf{18} & \textbf{20} & \textbf{23} \\
\cmidrule(lr){3-8}
\multicolumn{2}{l}{\textbf{HE-NAC CBs}
  $(J_\mathrm{CB}, J_\mathrm{MB}, J_\mathrm{HB})$}
&
\multicolumn{1}{c}{$(16,1,0)$} &
\multicolumn{1}{c}{$(16,2,0)$} &
\multicolumn{1}{c}{$(16,3,0)$} &
\multicolumn{1}{c}{$(16,4,0)$} &
\multicolumn{1}{c}{$(16,4,1)$} &
\multicolumn{1}{c}{$(16,4,2)$} \\
\multicolumn{2}{l}{\rebuttalvtwo{\textbf{DAC CBs}
  $(J)$}}
&
\multicolumn{1}{c}{\rebuttalvtwo{18}} &
\multicolumn{1}{c}{\rebuttalvtwo{20}} &
\multicolumn{1}{c}{\rebuttalvtwo{22}} &
\multicolumn{1}{c}{\rebuttalvtwo{24}} &
\multicolumn{1}{c}{\rebuttalvtwo{27}} &
\multicolumn{1}{c}{\rebuttalvtwo{31}} \\
\midrule
\textbf{Band} & \textbf{Model} & 
\multicolumn{6}{>{\columncolor{gray!20}}c}{\textbf{SI-SDR (dB)}} \\
\midrule

\multirow{2}{*}{0-3~kHz}
    & DAC    & \bfseries 15.8 & \bfseries 16.6 & \bfseries 17.3 & \bfseries 18.0 & \bfseries 19.0 & \bfseries 20.6 \\
    & HE-NAC & 15.4 & 15.4 & 15.4 & 15.4 & 15.4 & 15.4 \\
\addlinespace[0.4em]
\arrayrulecolor{black!15}\midrule
\arrayrulecolor{black}

\multirow{2}{*}{3-6~kHz}
    & DAC    & -5.2 & -3.5 & -2.3 & -1.4 & -0.3 &  0.7 \\
    & HE-NAC & \bfseries -2.8 & \bfseries 0.3 & \bfseries 2.3 & \bfseries 3.7 & \bfseries 3.7 & \bfseries 3.7 \\
\addlinespace[0.4em]
\arrayrulecolor{black!15}\midrule
\arrayrulecolor{black}

\multirow{2}{*}{6-16~kHz}
    & DAC    & \bfseries -36.3 & \bfseries -34.7 & \bfseries -33.60 & \bfseries -32.6 & -31.7 & -30.7 \\
    & HE-NAC & -39.1 & -37.8 & -36.2 & -35.4 & \bfseries -8.8 & \bfseries -5.5 \\
\addlinespace[0.4em]
\arrayrulecolor{black!15}\midrule
\arrayrulecolor{black}

\multirow{2}{*}{Full-band}
    & DAC    & \bfseries 12.6 & \bfseries 13.2 & \bfseries 13.7 & \bfseries 14.2 & \bfseries 14.9 & \bfseries 15.7 \\
    & HE-NAC & 12.4 & 12.7 & 12.9 & 13.0 & 13.6 & 13.7 \\
\midrule

 &  &
\multicolumn{6}{>{\columncolor{gray!20}}c}{\textbf{GOMPSNR (dB)}}\\
\midrule

\multirow{2}{*}{0-3~kHz}
    & \rebuttal{DAC}
    & \rebuttal{\bfseries 10.73}
    & \rebuttal{\bfseries 11.14}
    & \rebuttal{\bfseries 11.54}
    & \rebuttal{\bfseries 11.95}
    & \rebuttal{\bfseries 12.51}
    & \rebuttal{\bfseries 13.29} \\
    & \rebuttal{HE-NAC}
    & \rebuttal{10.43}
    & \rebuttal{10.43}
    & \rebuttal{10.43}
    & \rebuttal{10.43}
    & \rebuttal{10.43}
    & \rebuttal{10.43} \\
\addlinespace[0.4em]
\arrayrulecolor{black!15}\midrule
\arrayrulecolor{black}

\multirow{2}{*}{3-6~kHz}
    & \rebuttal{DAC}
    & \rebuttal{1.47}
    & \rebuttal{1.84}
    & \rebuttal{2.17}
    & \rebuttal{2.47}
    & \rebuttal{2.88}
    & \rebuttal{3.33} \\
    & \rebuttal{HE-NAC}
    & \rebuttal{\bfseries 1.86}
    & \rebuttal{\bfseries 2.81}
    & \rebuttal{\bfseries 3.58}
    & \rebuttal{\bfseries 4.25}
    & \rebuttal{\bfseries 4.16}
    & \rebuttal{\bfseries 4.16} \\
\addlinespace[0.4em]
\arrayrulecolor{black!15}\midrule
\arrayrulecolor{black}

\multirow{2}{*}{6-16~kHz}
    & \rebuttal{DAC}
    & \rebuttal{\bfseries 0.12}
    & \rebuttal{\bfseries 0.12}
    & \rebuttal{\bfseries 0.13}
    & \rebuttal{\bfseries 0.14}
    & \rebuttal{\bfseries 0.15}
    & \rebuttal{0.17} \\
    & \rebuttal{HE-NAC}
    & \rebuttal{-0.20}
    & \rebuttal{-0.15}
    & \rebuttal{-0.12}
    & \rebuttal{-0.10}
    & \rebuttal{-0.30}
    & \rebuttal{\bfseries 0.27} \\
\addlinespace[0.4em]
\arrayrulecolor{black!15}\midrule
\arrayrulecolor{black}

\multirow{2}{*}{Full-band}
    & \rebuttal{DAC}
    & \rebuttal{\bfseries 9.78}
    & \rebuttal{\bfseries 10.16}
    & \rebuttal{\bfseries 10.52}
    & \rebuttal{\bfseries 10.88}
    & \rebuttal{\bfseries 11.37}
    & \rebuttal{\bfseries 12.02} \\
    & \rebuttal{HE-NAC}
    & \rebuttal{9.54}
    & \rebuttal{9.64}
    & \rebuttal{9.71}
    & \rebuttal{9.77}
    & \rebuttal{9.86}
    & \rebuttal{9.90} \\

\midrule

 &  & 
\multicolumn{6}{>{\columncolor{gray!20}}c}{\textbf{VISQOL}} \\
\midrule
\addlinespace[0.3em]

\multirow{2}{*}{Full-band}
& DAC
  & \multicolumn{1}{c}{\num[round-precision=2]{4.26}}
  & \multicolumn{1}{c}{\bfseries \num[round-precision=2]{4.29}}
  & \multicolumn{1}{c}{\bfseries \num[round-precision=2]{4.30}}
  & \multicolumn{1}{c}{\bfseries \num[round-precision=2]{4.32}}
  & \multicolumn{1}{c}{\num[round-precision=2]{4.33}}
  & \multicolumn{1}{c}{\num[round-precision=2]{4.35}} \\
& HE-NAC
  & \multicolumn{1}{c}{\bfseries \num[round-precision=2]{4.27}}
  & \multicolumn{1}{c}{\bfseries \num[round-precision=2]{4.29}}
  & \multicolumn{1}{c}{\bfseries \num[round-precision=2]{4.30}}
  & \multicolumn{1}{c}{ \num[round-precision=2]{4.31}}
  & \multicolumn{1}{c}{\bfseries \num[round-precision=2]{4.36}}
  & \multicolumn{1}{c}{\bfseries \num[round-precision=2]{4.40}} \\

\bottomrule
\end{tabular}
}
\end{table}

\begin{figure}
    \centering
        \includegraphics[width=1.0\linewidth]{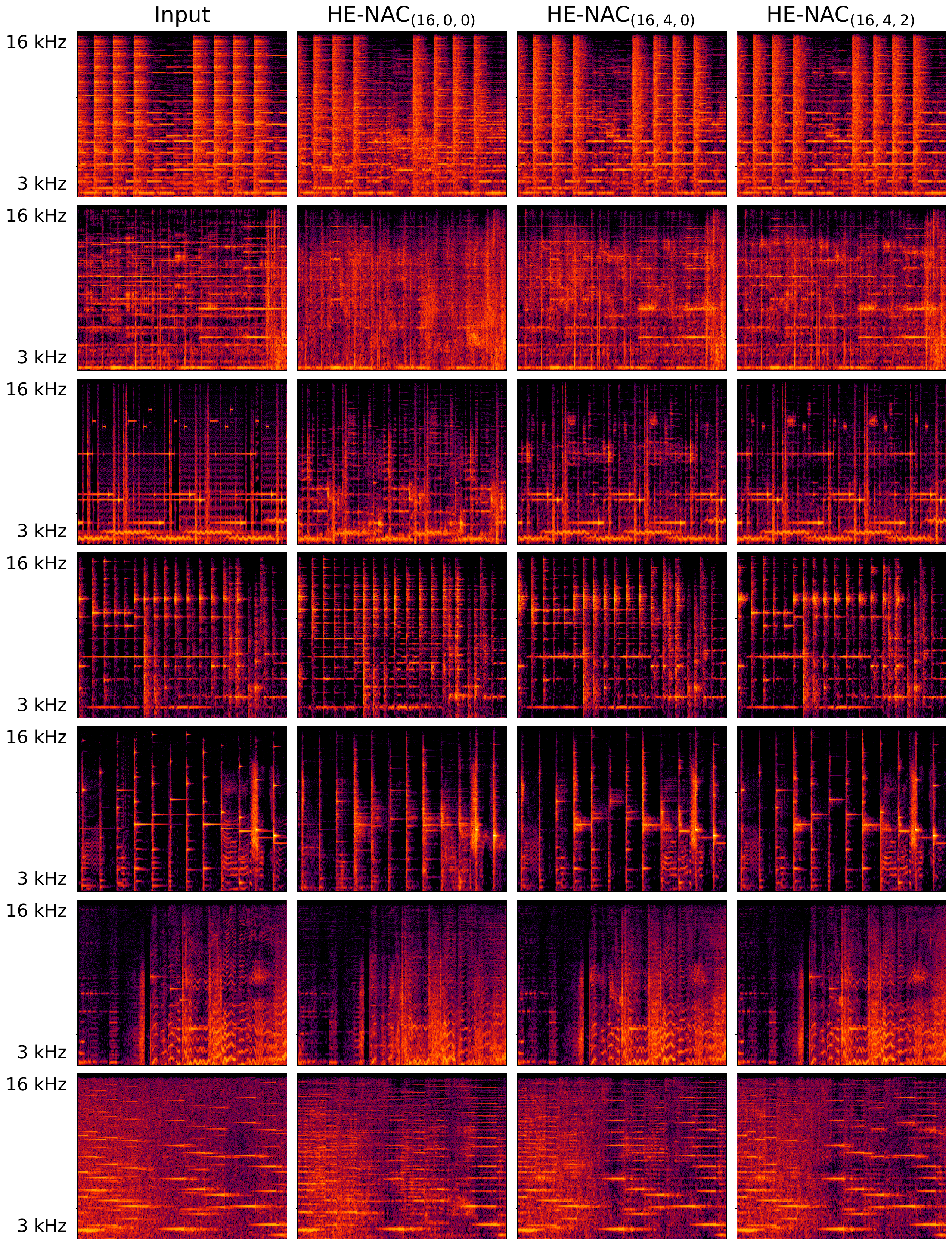}
    \caption{Comparison in the time-frequency domain between various HE-NAC configurations.}
    \label{fig:ablation}
\end{figure}

\rebuttal{To further evaluate the different information streams in HE-NAC, we complement Fig.~\ref{fig:distortion_rate} with Table~\ref{tab:sisdr_visqol_results}, which reports SI-SDR, VISQOL, and the recently proposed GOMPSNR metric~\cite{dai2026gompsnr} across the effective bitrate range. While SI-SDR primarily measures waveform fidelity, GOMPSNR has recently been proposed to better account for phase-related distortions in generative neural audio models, providing an additional perspective on reconstruction quality.}

Similarly to its STFT counterpart (see Fig.~\ref{fig:distortion_rate}), SI-SDR returns advantageous numbers in the CB range (0-3~kHz) for the DAC baseline while being outperformed by HE-NAC in the MB range when the MB codes start to kick in, as well as in the HB range with the help of HB codes. Note that these improvements are not directly reflected in the full-band numbers, both in SI-SDR and VISQOL, although our subjective tests disclosed the perceptual relevance of the MB and HB ranges successfully.

\rebuttal{GOMPSNR exhibits a similar trend in the CB and MB regions, consistently favoring DAC in the CB range and HE-NAC in the MB range across all evaluated bitrates. However, unlike SI-SDR, the HB and full-band GOMPSNR scores slightly favor DAC. Together with the statistically significant preference for HE-NAC observed in the MUSHRA listening tests, these results suggest that different objective metrics emphasize different aspects of reconstruction quality.}

Fig.~\ref{fig:ablation} displays more concrete visuals of this upper-band reconstruction gain in the form of spectrograms (3-16kHz). As more band-specific decoders are added, namely MB (third column) and HB (fourth column), the upper-band reconstruction quality becomes increasingly more accurate.

\begin{figure}
\centering
    \includegraphics[width=\linewidth]{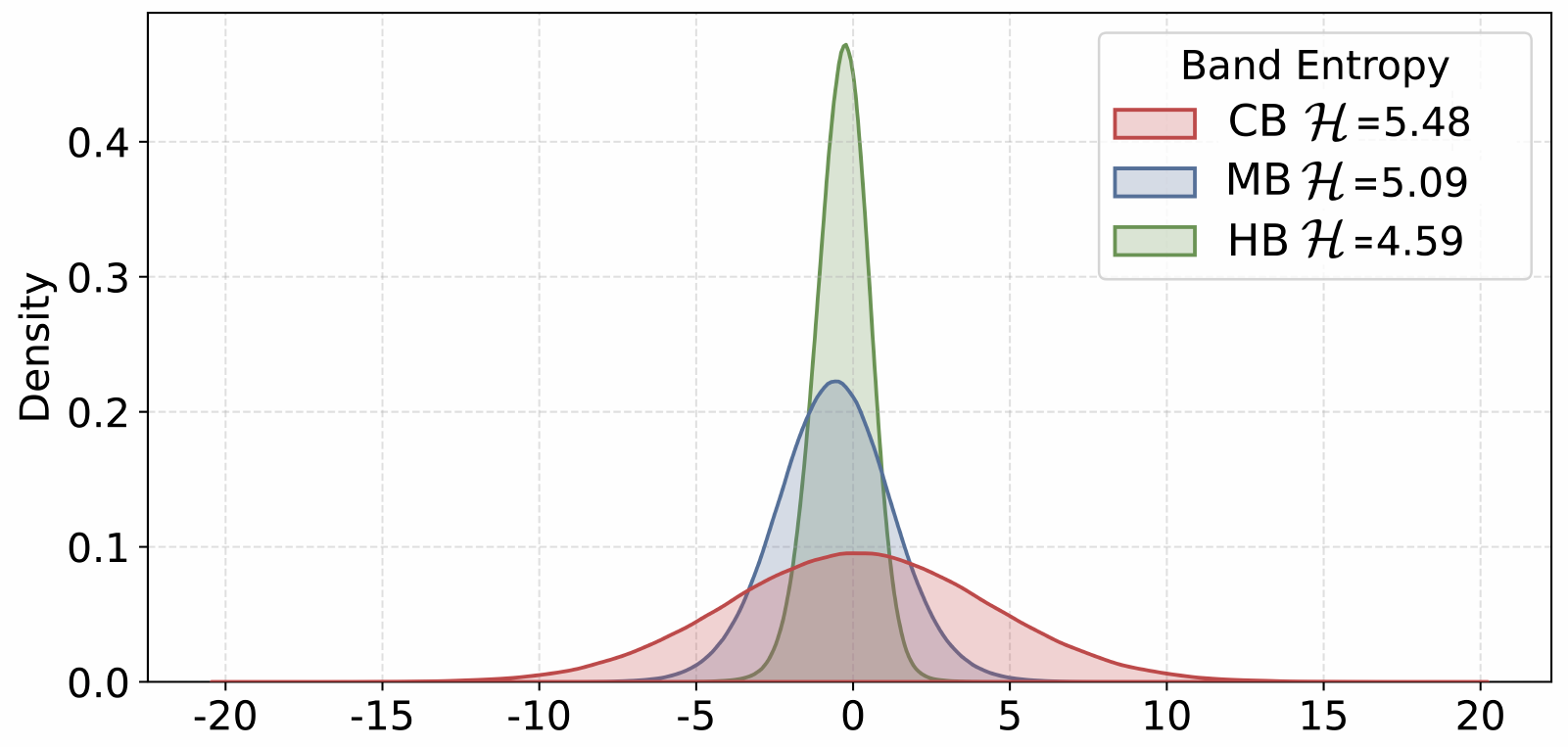}
  \caption{The distribution of per-band feature vectors. The latent features per bottleneck are aggregated and concatenated over the audio files, and histogram are drawn from them. Their respective Shannon entropy is also shown for each of the distributions.}
  \label{fig:hist}    
\end{figure}

Lastly, we examine the distributions of the subband-specific code spaces. From our full model, $\text{HE-NAC}_{(16, 4, 2)}$, Fig.~\ref{fig:hist} shows latent histograms and empirical entropy. CB (red) exhibits the most long-tail distribution with a high entropy of $5.48$, while the entropy gradually decreases in MB (blue) and HB (green). These sample distributions and their associated entropy values showcase the residual-like characteristics of the higher-band information streams, which HE-NAC enforces using the multi-stage training algorithm (see Algo.~\ref{alg:he-nac}) and the hybrid architecture for blind upsampling and residual learning. These patterns further validate the structured complexity of audio signals across bands and motivate our band-specific quantization approach.

\subsection{Inference Complexity}

\rebuttal{To assess the computational cost introduced by the proposed HE-NAC architecture, Table~\ref{tab:inference_complexity} compares parameter-matched DAC and HE-NAC variants in terms of active parameters, real-time factor (RTF), and normalized computational complexity (GFLOPs per second of audio). All runtime measurements were performed in FP32 on a single NVIDIA A100 40\,GB GPU. Across all configurations, both codecs operate substantially faster than real time (RTF $\ll 1$). The HE-NAC$_{(16,4,0)}$ variant exhibits nearly identical inference speed to the parameter-matched DAC baseline while requiring approximately 9\% fewer GFLOPs. At the highest operating point, HE-NAC$_{(16,4,2)}$ incurs a moderate increase in computational cost (0.0165 vs.\ 0.0112 RTF and 491.7 vs.\ 426.7 GFLOPs/audio s), reflecting the additional computation associated with the dedicated high-band residual coding. Nevertheless, the proposed architecture maintains faster-than-real-time offline inference while delivering the perceptual quality improvements demonstrated in the previous sections; neither codec currently supports streaming inference.}

\begin{table}
\color{black}
\caption{Inference complexity of parameter-matched DAC and HE-NAC variants.
RTF is the ratio of inference time to audio duration; values below one indicate
faster-than-real-time processing. GFLOPs are normalized per second of 32-kHz
mono audio using $1$ MAC $=2$ FLOPs. Measurements were obtained with batch
size one over 39 ten-second clips (390 seconds total), excluding file I/O and
metric computation. All models were evaluated offline and do not inherently
support streaming inference.}
\centering
\begin{tabular}{l c c c}
\toprule
\rowcolor{gray!20}
\textbf{Model} &
\makecell{\textbf{Active}\\\textbf{\#Params}} &
\textbf{RTF} $\downarrow$ &
\makecell{\textbf{GFLOPs /}\\\textbf{audio s} $\downarrow$} \\
\toprule
DAC$_{(31)}$ & 85M & 0.0066 & 149.9 \\

\midrule
DAC$_{(31)}$      & 115M & 0.0096 & 337.2 \\
HE-NAC$_{(16,4,0)}$ & 127M & 0.0095 & 308.1 \\

\midrule
DAC$_{(31)}$     & 220M & 0.0112 & 426.7 \\
HE-NAC$_{(16,4,2)}$ & 220M & 0.0165 & 491.7 \\
\bottomrule
\end{tabular}
\label{tab:inference_complexity}
\end{table}

\subsection{Sound Examples, Source Codes, and Checkpoints}

We release the code and trained checkpoints for reproducibility\footnote{\url{https://minjekim.com/research-projects/he-nac}}. We also provide a collection of audio examples corresponding to twelve excerpts from the test set. All files are organized for straightforward cross-system comparison. Detailed descriptions of the file structure and naming conventions are provided in the supplementary material.

\section{Conclusion}
We presented a novel high-fidelity neural audio coding system, HE-NAC. It is a multi-band neural audio codec capable of accurately reconstructing high-frequency components. Our approach disclosed the residual nature of upper bands reconstructions by leveraging a multi-stage residual learning algorithm. Each stage of our model exploits latent features from previous bands for blind upsampling, as well as the band-specific codes that compensate for the blind upsampling scheme. Hence, it provides a balanced way of harmonizing bitrate-free blind upsampling and multi-band coding. Ablation studies confirm the benefit of combining band-specific and cross-band cues for preserving high-frequency details typically lost in single-bottleneck models. Objective and subjective results show that HE-NAC's multi-band bit allocation provides advanced sound quality compared to the DAC baseline and the commercial implementation of HE-AAC v1.

\bibliographystyle{IEEEtran}
\bibliography{mjkim}  %

\end{document}